\documentclass[journal, romanappendices]{IEEEtran}
\usepackage{amsmath,amsfonts}
\usepackage{algorithmic}
\usepackage{algorithm}
\usepackage{array}
\usepackage[caption=false,font=normalsize,labelfont=sf,textfont=sf]{subfig}
\usepackage{textcomp}
\usepackage{cite}
\usepackage{stfloats}
\usepackage{url}
\usepackage{verbatim}
\usepackage{graphicx}
\usepackage{multirow}
\usepackage{ragged2e}
\usepackage{booktabs}
\usepackage{threeparttable}
\usepackage{soul}
\usepackage{tabularx}
\usepackage{color,xcolor}
\soulregister{\cite}7
\soulregister{\citep}7 
\soulregister{\citet}7 
\soulregister{\ref}7 
\usepackage{CJK}
\usepackage{makecell}
\allowdisplaybreaks[2]
\usepackage{indentfirst}
\usepackage{float}
\usepackage[amsthm,amsmath,thmmarks]{ntheorem} 
\usepackage[numbers,sort&compress]{natbib}
\def\BibTeX{{\rm B\kern-.05em{\sc i\kern-.025em b}\kern-.08em
    T\kern-.1667em\lower.7ex\hbox{E}\kern-.125emX}}
\usepackage{balance}
\begin{document}
\title{XCloud-VIP: Virtual Peak Enables Highly Accelerated NMR Spectroscopy and Faithful Quantitative Measures}
\author{Di Guo,  Zhangren Tu, Yi Guo, Yirong Zhou, Jian Wang, Zi Wang, Tianyu Qiu, Min Xiao, Yinran Chen, Liubin Feng, Yuqing Huang, Donghai Lin, Qing Hong, Amir Goldbourt, Meijin Lin and Xiaobo Qu*
\thanks{This work was supported by National Natural Science Foundation of China (62371410, 62122064, 61971361, 62331021), Natural Science Foundation of Fujian Province of China (2021J011184), President Fund of Xiamen University (20720220063), Nanqiang Outstanding Talents Program of Xiamen University.} 
\thanks{Di Guo and Yi Guo are with the School of Computer and Information Engineering, Fujian Engineering Research Center for Medical Data Mining and Application, Xiamen University of Technology, Xiamen 361024, China.}
\thanks{Zhangren Tu, Yirong Zhou, Jian Wang, Zi Wang, Tianyu Qiu, Yuqing Huang, and Xiaobo Qu are with the Department of Electronic Science, Biomedical Intelligent Cloud R\&D Center, Fujian Provincial Key Laboratory of Plasma and Magnetic Resonance, Xiamen University, Xiamen 361102, China.} 
\thanks{Min Xiao is with the School of Opto-Electronic and Communication Engineering, Xiamen University of Technology, Xiamen 361024, China.}
\thanks{Yinran Chen is with the Department of Computer Science, School of Information, Xiamen University, Xiamen 361102, China.}
\thanks{Liubin Feng and Donghai Lin are with the College of Chemistry and Chemical Engineering, Key Laboratory for Chemical Biology of Fujian Province, High-field NMR Center, Xiamen University, Xiamen 361005, China.} 
\thanks{Qing Hong is with the China Mobile Group, Xiamen 361005, China.}
\thanks{Amir Goldbourt is with the School of Chemistry, Tel Aviv University, Tel Aviv 6997801, Israel.}
\thanks{Meijin Lin is with the Department of Applied Marine Physics and Engineering, College of Ocean and Earth Sciences, Xiamen University, Xiamen 361102, China.}
\thanks{*Corresponding author: quxiaobo@xmu.edu.cn}}

\markboth{Journal of \LaTeX\ Class Files}%
{How to Use the IEEEtran \LaTeX \ Templates}

\maketitle

\begin{abstract}
Nuclear Magnetic Resonance (NMR) spectroscopy is an important bio-engineering tool to determine the metabolic concentrations, molecule structures and so on. The data acquisition time, however, is very long in multi-dimensional NMR. To accelerate data acquisition, non-uniformly sampling is an effective way but may encounter severe spectral distortions and unfaithful quantitative measures when the acceleration factor is high. By modelling the acquired signal as the superimposed exponentials, we proposed a virtual peak (VIP) approach to self-adapt the prior spectral information, such as the resonance frequency and peak lineshape, and then feed these information into the reconstruction. The proposed method is further implemented with cloud computing to facilitate online, open, and easy access. Results on simulated and experimental data demonstrate that, compared with the low-rank Hankel matrix method, the new approach reconstructs high-fidelity NMR spectra from highly undersampled data and achieves more accurate quantification. The maximum quantitative errors of distances between nuclear pairs and concentrations of metabolites in mixtures have been reduced by 61.1\% and 57.7\%, respectively.
\end{abstract}

\begin{IEEEkeywords}
machine learning, nuclear magnetic resonance spectroscopy, fast sampling, Hankel matrix.
\end{IEEEkeywords}

\section{Introduction}
\IEEEPARstart{N}{uclear} Magnetic Resonance (NMR) spectroscopy is an important analytical tool in medicine, biology and chemistry. Multidimensional NMR provides fruitful information, such as spin-spin coupling and molecular structures, but the data acquisition time increases significantly with spectral resolution and dimensions \cite{1,2,3}.

Accelerating data acquisition is one of the major developments in modern NMR \cite{1,2,3,4,5,6,7,8,9,10,11,12,13}. Without changing the equipment, Non-Uniform Sampling (NUS) allows acquiring fewer data points and spectrum reconstruction with proper signal priors \cite{1,5,6,7,8,10,14,15,16}, such as spectrum sparsity in compressed sensing \cite{5,6,7,8}, the minimal number of peaks in low-rank \cite{3,10,13,16} and deep learning \cite{11,12}. Although these methods are powerful for spectrum reconstructions, they still suffer from spectral distortion under a high acceleration factor. 

Why is there distortion? A reconstruction means implicitly restoring lots of peak information, including amplitude, frequency, phase and damping factor. The amplitude determines the intensity of a peak and the latter three factors define the location and lineshape of a peak. Thus, these four factors contribute to the four degrees of freedom of a spectrum. Hence, if a reconstruction approach is not designed properly, the degree of freedom may be too high to obtain a good spectrum.   

To reduce the spectrum distortion, a possible solution is to reduce the degree of freedom (DOF) by mining the prior information as much as possible. One fundamental approach is to model the acquired signal, i.e. Free Induction Decay (FID), as a linear superposition of modulated exponential functions \cite{13,16,17,18,19,20,21}. This model is convenient to incorporate priors since each exponential function corresponds to one spectral peak. However, mining each peak from the spectrum is not trivial since the spectrum is a superimposed signal. 

In this paper, we first get a reference spectrum by reconstructing the undersampled FID, and then employ the classic Hankel singular value decomposition (SVD) \cite{3,13,18,22} to extract each VIrtual Peak (VIP), and finally feed this information into a proposed signal reconstruction model.

It is worth noting that modeling the acquired signal as exponentials was also investigated in other computational imaging fields, such as magnetic resonance imaging (MRI) \cite{18,23,24,25,26,27,28,40}, radar imaging \cite{29} and geoscience imaging \cite{30}. Thus, our work is not only valuable for NMR, but also for other areas where prior information of exponentional modeling is introduced.

 The structure of the paper is as follows: Section II presents a mathematical model and derives a numerical algorithm. Section III describes the experimental results. Section IV is the discussion and Section V concludes this work.

\begin{figure*}[!t]
\centering
\includegraphics[width=6.8in]{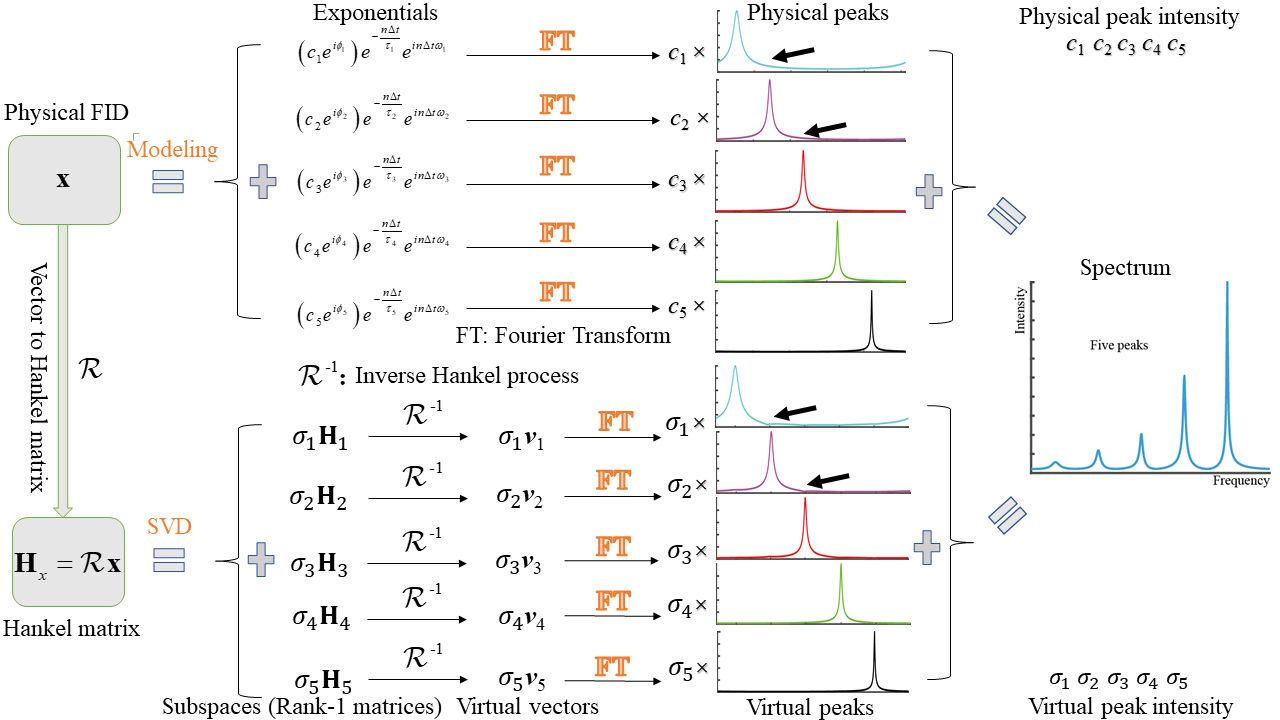}
\caption{The schematic of learning virtual peaks (VIPs) from a FID with 5 peaks. Note: The notation   means the linear combination of these virtual (or physical) peaks will compose the spectrum and the combination weight is called the virtual (or physical) peak intensity.}
\label{fig1}
\end{figure*}

\section{METHOD}
In this work, noise is not considered in the forward problem of signal modeling (Sections II.A and II. B) but considered in solving the inverse problem, including the reconstruction model and algorithm (Sections II. C and II. D)
\subsection{Modelling of Virtual Peaks}
Our method is based on a basic modelling of FID signal as the sum of exponentials and a property that the number of peaks is equal to the rank of the Hankel matrix converted from the FID \cite{3,11,13,18,22,41}.

 A typical noise-free FID can be described as the sum of exponentially decaying sinusoids according to
\begin{equation}
\label{deqn_ex1}
{{x}_{n}}\text{=}\sum\limits_{p=1}^{P}{{{c}_{p}}{{e}^{i{{\phi }_{p}}}}{{e}^{in\Delta t{{\omega }_{p}}}}{{e}^{-\frac{n\Delta t}{{{\tau }_{p}}}}}},
\end{equation}
where ${{x}_{n}}$ denotes the ${{n}^{th}}$  element of the FID $\mathbf{x}\text{=}{{\left[ {{x}_{0}},{{x}_{1}},\cdots ,{{x}_{2N}} \right]}^{T}}$, while ${{c}_{p}}$, ${{\phi }_{p}}$, ${{\omega }_{p}}$ and ${{\tau }_{p}}$ denote the amplitude, phase, resonance frequency and $T_{2}^{*}$ of the ${{p}^{th}}$ exponential (or peak), respectively. The total number of exponentials (or peaks) is $P$. The $\Delta t$ is the sampling interval. The representative spectra (non-equidistant resonances, different linewidths and line shapes, and overlapping) are used in the simulation. The reconstruction experiment of triplet-peak spectra and the parameter settings of synthesizing the spectra are reported in Supplements S1 and S9, respectively. We further discussed the reconstruction results of non-ideal Lorentzian lineshaped peaks, as detailed in Discussion.

Let $\mathsf{\mathcal{R}} $ denote a Hankel operator converting  $\mathbf{x} $ into a Hankel matrix $\mathsf{\mathcal{R}}\mathbf{x}\in {{\mathbb{C}}^{N\times N}}$, and the SVD of $\mathsf{\mathcal{R}}\mathbf{x} $ \cite{3,11,13,18,22} is
\begin{equation}
\label{deqn_ex2}
\mathsf{\mathcal{R}}\mathbf{x}\text{=}\mathbf{A\Lambda }{{\mathbf{B}}^{H}},
\end{equation}
where \textbf{A} (or \textbf{B}) is the left (or right) signal space and $\mathbf{\Lambda }$ is a diagonal matrix called the singular value matrix, the superscript \textsl{H} denotes the Hermitian transpose. The rank of Hankel matrix is defined as the number of non-zero singular values in $\mathbf{\Lambda }$, whose diagonal entries, i.e. singular values, are commonly stored in descending order. For a given spectrum with \textsl{P} peaks, the rank of the Hankel matrix is \textsl{P} \cite{3,13}.

A Hankel matrix with rank \textsl{P} can be linearly combined by \textsl{P} Rank-1 matrices ${{\left\{ {{\mathbf{H}}_{p}}\in {{\mathbb{C}}^{N\times N}} \right\}}_{p=1,2,\cdots ,P}} $ as
\begin{equation}
\label{deqn_ex3}
\mathsf{\mathcal{R}}\mathbf{x}=\sum\limits_{p=1}^{P}{{{\sigma }_{p}}{{\mathbf{H}}_{p}}},    
\end{equation}
where ${{\sigma }_{p}}$ is the ${{p }^{th}}$ singular value stored in $\mathbf{\Lambda}$. For each ${{\mathbf{H}}_{p}}$, an inverse Hankel process is defined as ${{\mathsf{\mathcal{R}}}^{-1}}$ according to
\begin{equation}
\label{deqn_ex4}
{{\mathsf{\mathcal{R}}}^{-1}}:{{\mathbb{C}}^{N\times N}}\to {{\mathbb{C}}^{2N+1}},
\end{equation}
by performing the inverse Hankel process using an operator ${{\mathsf{\mathcal{R}}}^{-1}}$ to convert a subspace matrix ${{\mathbf{H}}_{p}}$ to a vector ${{\mathbf{v}}_{p}}\text{=}{{\mathsf{\mathcal{R}}}^{-1}}{{\mathbf{H}}_{p}}$ via averaging anti-diagonal entries of ${{\mathbf{H}}_{p}}$. ${{\mathbf{v}}_{p}}$ is also called the virtual vector in Figure \ref{fig1}. Then, a VIP ${{\mathbf{s}}_{p}}$ is obtained by performing the Fourier transform $\mathsf{\mathcal{F}}$ on each vector ${{\mathbf{v}}_{p}}$ as
\begin{equation}
\label{deqn_ex5}
{{\mathbf{s}}_{p}}=\mathsf{\mathcal{F}}{{\mathbf{v}}_{p}}.
\end{equation}

Accordingly, the ${{\sigma }_{p}}$ is also the virtual intensity for the ${{p }^{th}}$ peak as shown in Figure \ref{fig1}.

Another decomposition form of the Hankel matrix $\mathsf{\mathcal{R}}\mathbf{x} $ is known as the Vandermonde decomposition\cite{10}, which explicitly preserves exponential information (physical peak). Its form is as follows:

\begin{equation}
\mathcal{R} \mathbf{x}=\boldsymbol{E} \boldsymbol{\Sigma} \boldsymbol{E}^T
\end{equation}
where$$
\boldsymbol{E} = \left [\begin{matrix}
   z_{1}^{0} & z_{2}^{0} & \cdots  & z_{P}^{0}  \\
   z_{1}^{1} & z_{2}^{1} & \cdots  & z_{P}^{1}  \\
   \vdots  & \vdots  & \ddots  & \vdots   \\
   z_{1}^{N-1} & z_{2}^{N-1} & \cdots  & z_{P}^{N-1}  \\
\end{matrix} \right] \in {{\mathbb{C}}^{N\times P}}$$
are the Vandermonde matrix that saves the physical peak in each column, $ {{z}_{p}}={{e}^{(i{{w}_{p}}-\frac{1}{{{\tau }_{p}}})}}$,  and $\boldsymbol{\Sigma}=\operatorname{diag}\left\{c_p e^{i \phi_p}\right\} \in \mathbb{C}^{P \times P}, p=1, \ldots, P $ is a diagonal matrix, the superscript $T$ denotes the Hermitian transpose.

The whole process of VIP extraction is shown in Figure \ref{fig1}. 

\begin{figure}[!t]
\centering
\includegraphics[width=3.5in]{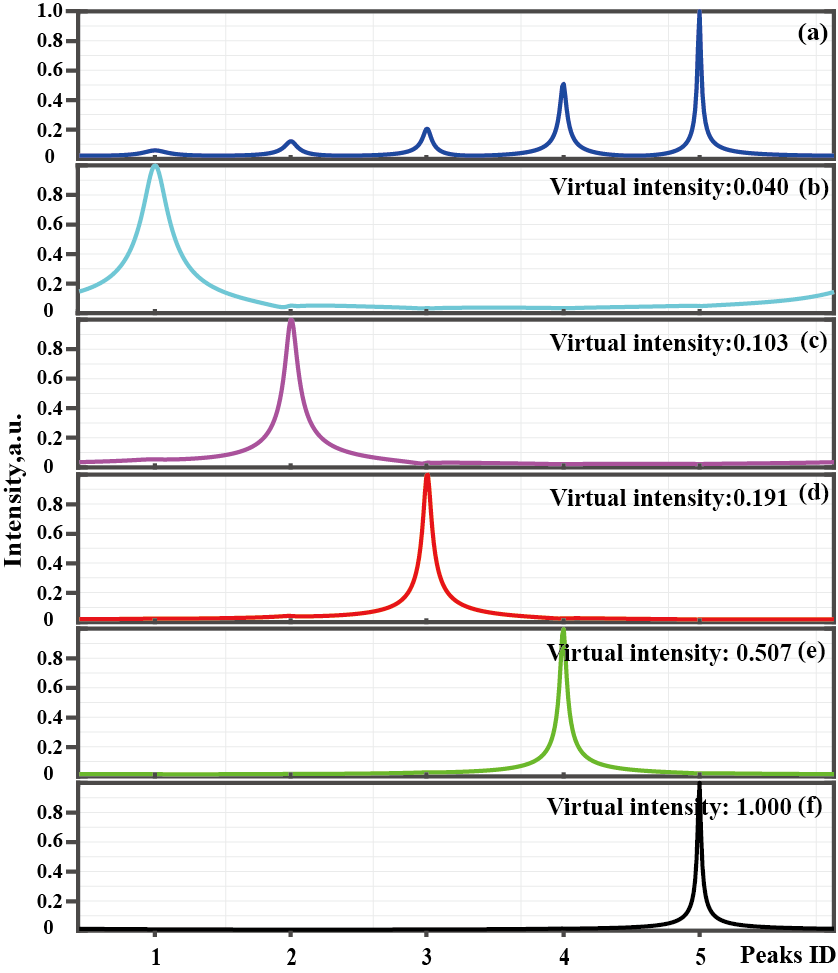}
\caption{Virtual peaks of a spectrum. (a) the original noise-free spectrum with 5 physical peaks, (b)-(f) virtual peaks that satisfies (a) = 0.040×(b) + 0.103×(c) + 0.191×(d) + 0.507×(e) + 1.000×(f). \hl{}} 
\label{fig2}
\end{figure}

However, as the FID is a superimposed signal of these peaks, mining each  peak is not easy. In general, one-to-one correspondence between the VIP and the physical peak is hard since the SVD enforces the orthogonality of each column (and row) in the rank-1 matrix. In practical NMR reconstructions, only FID, i.e. the linear combination of multiple peaks, is available. The purpose of defining VIPs is to present the great potential of subspaces (or SVD) to incorporate the prior knowledge of a reference spectrum.  In this noise-free example, some details of peaks that are marked by arrow may be slightly different even though these VIPs are empirically observed to carry most spectral information, such as the resonance frequency and the lineshape of each peak (Figure \ref{fig2}).The linear combination of the strongest 5 VIPs is close to the original physical spectrum, implying that information is not obviously lost in the VIP decomposition. However, noise in the FID does leak into these strong singular values and the true rank is not exactly equal to the \textsl{P} due to the non-zero singular values introduced by noise. This indicates that the VIP methods perform better for relatively high signal-to-noise ratio NMR.

\subsection{Self-adaptive Virtual Peaks}
How to learn reliable VIPs is important for reconstructions since the fully sampled FID is not available in NUS. Here, we suggest dividing VIPs into strong and weak peaks according to their virtual intensities. This process is relatively easy since the virtual intensity, i.e. the singular value, can be obtained with SVD on the Hankel matrix. In practice, we observe that strong VIPs are much more reliable than the weak ones.

\begin{figure}[!t]
\centering
\includegraphics[width=3.4in]{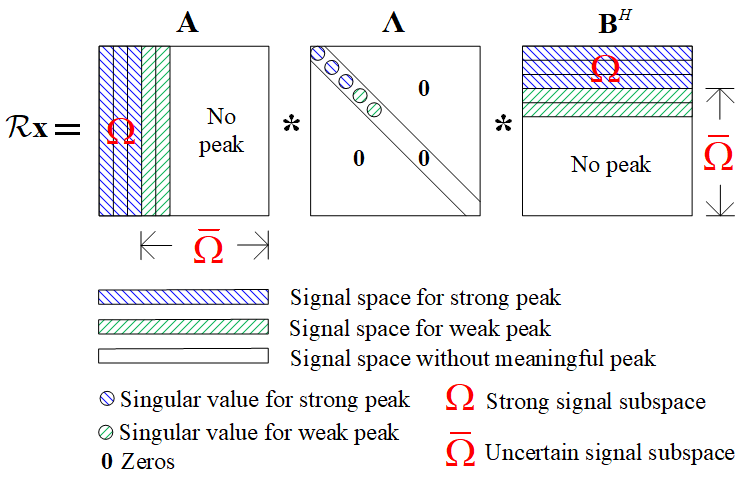}
\caption{Illustration on utilizing the reliable VIP from strong peaks.}
\label{fig3}
\end{figure}

As shown in Figure \ref{fig3}, strong signal subspace $\Omega $ denotes the prior VIP information of strong peaks, and uncertain signal subspace $\bar{\Omega }$ represents the rest of peaks. Then, $ \mathbf{A}\text{=}{{\mathbf{A}}_{\Omega }}\bigcup {{\mathbf{A}}_{{\bar{\Omega }}}} $ (and $ \mathbf{B}\text{=}{{\mathbf{B}}_{\Omega }}\bigcup {{\mathbf{B}}_{{\bar{\Omega }}}}$) where ${{\mathbf{A}}_{\Omega }}$ (and ${{\mathbf{B}}_{\Omega }}$) are the first $ p\left( p\le P \right)$ columns and ${{\mathbf{A}}_{{\bar{\Omega }}}}$ (and ${{\mathbf{B}}_{{\bar{\Omega }}}}$ ) are the rest $N-p$ columns of $\mathbf{A}$ (and $\mathbf{B}$). It should be noted that the first $p$ columns correspond to the subspace for the largest $p$ singular values stored in $\mathbf{\Lambda }$. Thus, the $p$ is a parameter that determines the number of prior strong peaks. The influence of this parameter on the reconstruction is detailed in Section III.G.

As the FID is undersampled, a reasonable solution is to utilize state-of-the-art reconstruction methods, such as compressed sensing \cite{31} or low-rank \cite{13}, to obtain a good reference spectrum. Surprisingly, in practice, we found that the proposed method is insensitive to initial reference if the VIP is updated several times (See Section IV). We simply choose the zero-filled spectrum as the initial reference to avoid using multiple algorithms.

\begin{figure}[!tbh]
\centering
\includegraphics[width=3.4in]{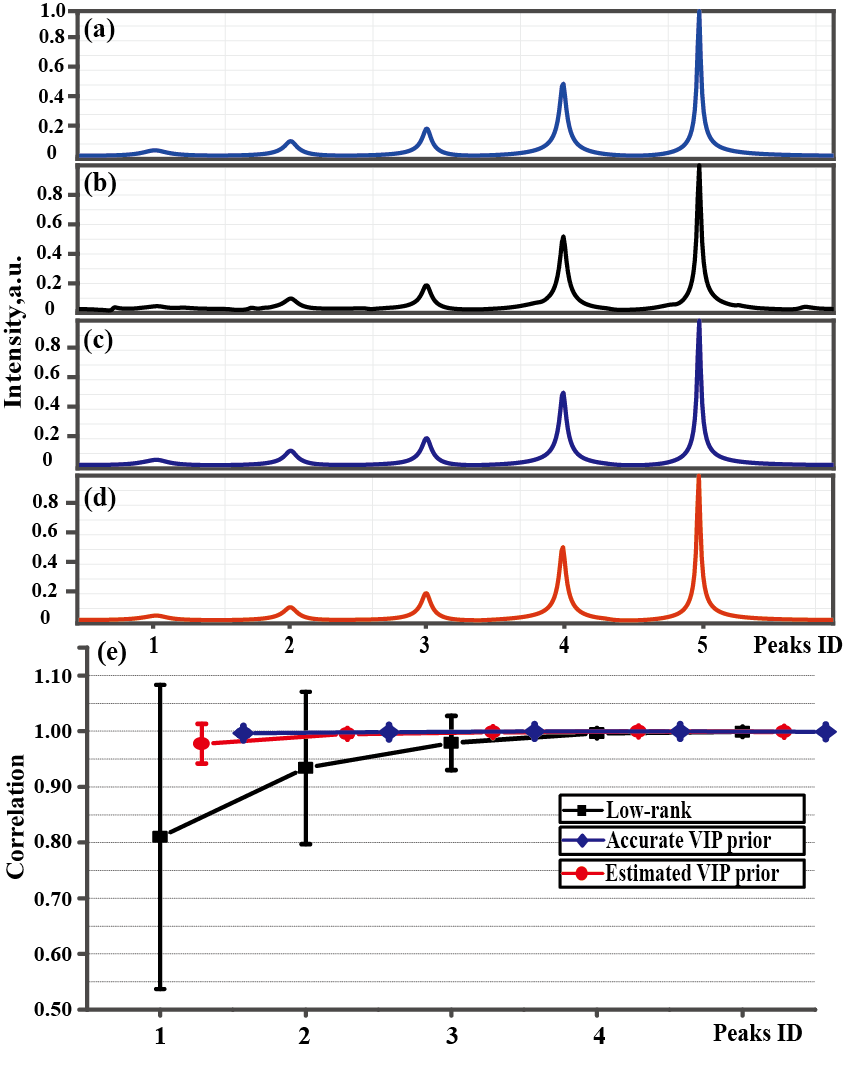}
\caption{Reconstruction of spectrum with 5 peaks and peak intensity correlations for the reconstructed spectra when the sampled data is 8\% of the fully sampled data. (a) is the original physical spectrum; (b) is reconstructed by low-rank; (c) is reconstructed with accurate prior of five VIPs (virtual intensity is not prior); (d) is reconstructed by the 3 most reliable, e.g. the highest three, VIPs; (e) is the peak intensity correlations of each peak. Note: The black, red and blue lines represent reconstructions by the low-rank, the proposed self-learning VIP method, and ideally accurate VIPs, respectively. The error bars are standard deviations of correlations over 100 sampling trials. A smaller bar indicates more robust reconstruction under different NUS trials. Note: Gaussian noise with zero mean and a standard deviation of 0.005 is added to the FID.}
\label{fig4}
\end{figure}

By introducing the VIP into the reconstruction model, the difficulty of solving the reconstruction problem will be reduced since fewer degrees of freedom for spectral peaks need to be estimated implicitly. Taking Figure \ref{fig4} as a noise toy example, it has an extremely high acceleration factor of 12.5. If this information taken by VIP is accurate, distorted peaks (peaks 1 and 2) can be reconstructed very faithfully, which implies that the VIP has a strong ability to obtain prior information. Even when choosing the zero-filled spectrum as the reference, by learning the three strongest VIPs from the reference then updating references and then learning the information again, the challenging low-intensity peaks will be restored very well (Figures \ref{fig4}(d) and \ref{fig4}(e)).

\subsection{Reconstruction Model of VIP}
The proposed model is defined to feed the VIP information $\Omega $ of strong peaks into reconstruction as follows:
\begin{equation}
\label{ex_6}  
\underset{\mathbf{x}}{\mathop{\text{min}}}\,{{\left\| \mathcal{R}\mathbf{x} \right\|}_{*}}-\operatorname{Tr}\left( \mathbf{A}_{\Omega }^{H}\mathcal{R}\mathbf{x}{{\mathbf{B}}_{\Omega }} \right)+\frac{\lambda }{2}\left\| \mathbf{y}-\mathsf{\mathcal{U}}\mathbf{x} \right\|_{2}^{2},
\end{equation}
where $\mathbf{x}\in {{\mathbb{C}}^{2N\text{+}1}}$ is the FID to be reconstructed, $\mathbf{y}\in {{\mathbb{C}}^{M}}$ are the acquired FID data, $\mathsf{\mathcal{U}}$ is an undersampling operator, the NUS rate is the percentage of the data obtained with NUS from the fully sampled data, i.e. $M/(2N+1)$, ${{\left\| \cdot  \right\|}_{2}}$ represents the ${{l}_{2}}$ norm and $\lambda$ is a parameter to balance the two terms, ${{\left\| \cdot  \right\|}_{*}}$  represents the nuclear norm (sum of singular values), $\operatorname{Tr}\left( \cdot  \right)$ is a trace function (sum of the main diagonal elements), the superscript \textsl{H} denotes the Hermitian transpose. Ideally, if the VIP of strong peaks is accurate,  $\mathbf{A}_{\Omega}^{H}\mathsf{\mathcal{R}}\mathbf{x}{{\mathbf{B}}_{\Omega }}$ will be a diagonal matrix of rank $P$ since both $\mathbf{A}_{\Omega }^{H}\mathbf{A}_{\Omega }^{{}}$ and $\mathbf{B}_{\Omega }^{H}\mathbf{B}_{\Omega }^{{}}$ will be the identity matrix. Thus, $ \operatorname{Tr}\left( \mathbf{A}_{\Omega}^{H}\mathsf{\mathcal{R}}\mathbf{x}{{\mathbf{B}}_{\Omega }} \right)$ sums up the largest $p$ singular values, i.e. virtual intensities. In other words, $ \operatorname{Tr}\left( \mathbf{A}_{\Omega}^{H}\mathsf{\mathcal{R}}\mathbf{x}{{\mathbf{B}}_{\Omega }} \right) $ defines the nuclear norm \cite{32}, which is considered as a convex measure for the number of  strong peaks \cite{13}. Therefore, the model tries to find the minimal number of weak peaks with the prior knowledge of VIP.

\begin{figure}[!ht]
\centering
\includegraphics[width=3.3in]{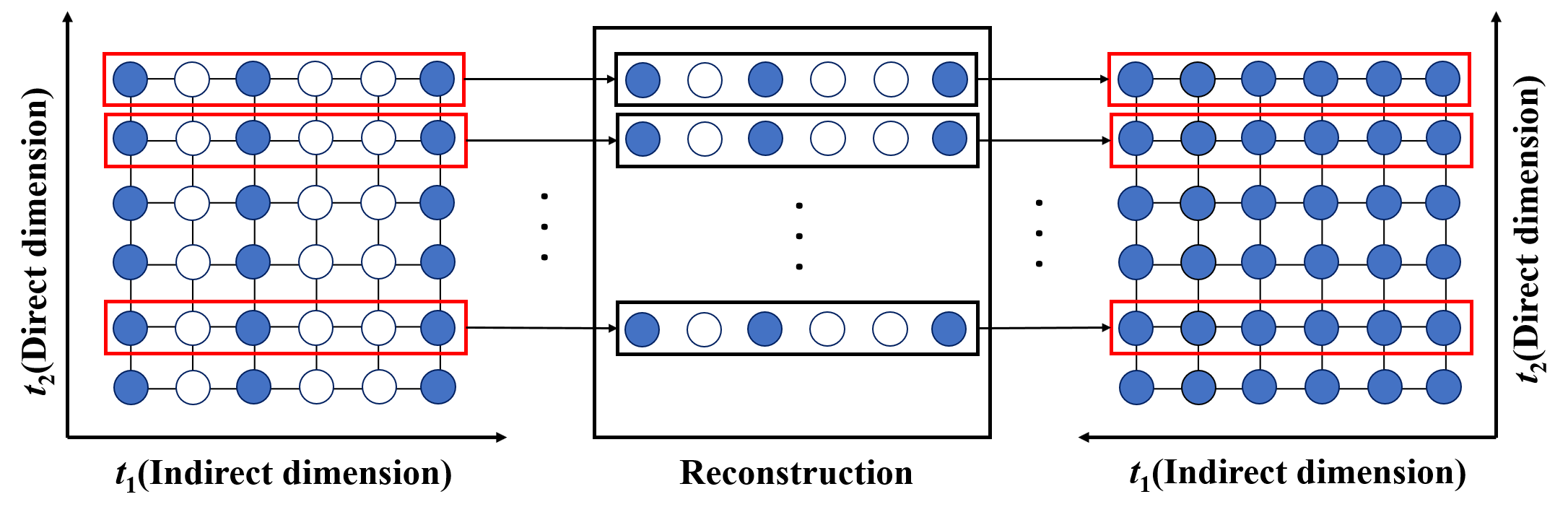}
\caption{The schematic diagram of reconstructing 2D NMR spectroscopy. Note: the blue and white points
represent the acquired data and non-acquired data, respectively.}
\label{fig05}
\end{figure}

According to NMR sampling theory, the acquisition of a 2D NMR spectrum is time-consuming because it takes a long time to sample in the indirect dimension (${{t}_{1}}$) due to the evolution of magnetization. The sampling in the direct dimension (${{t}_{2}}$)  is very fast. Therefore, 1D NUS is commonly applied to accelerate 2D NMR \cite{3}. The 2D NMR spectroscopy is generated by arranging the 1D spectra against ${{t}_{2}}$ (direct dimension) and the schematic diagram of reconstructing 2D NMR spectroscopy is shown in Figure \ref{fig05}.

This proposed model is inspired by the truncated nuclear norm regularization in general matrix completion \cite{33} but we solve a new reconstruction problem of recovery of missing data of exponential functions. Besides, we have provided a clear interpretation of prior VIP information in NMR reconstruction, verified the performance on the biological spectra, and implemented the algorithm on a cloud computing platform.

To further evaluate the performance of the proposed method,  different cases and different noise levels are reported in the  Sections III.B and III.F. Moreover, this paper summarizes the pros and cons of several state-of-the-art undersampled NMR spectrum reconstruction methods, including CS \cite{8} and DLNMR \cite{11} techniques (See detail in Disscussion). The results showed the proposed method outperforms CS \cite{8} and DLNMR \cite{11}.

\subsection{Numerical Algorithm}
For the NUS reconstruction, we suggest estimating the initial ${{\mathbf{A}}_{\Omega }}$ ( and ${{\mathbf{B}}_{\Omega }}$) from an initial solution ${{\mathbf{x}}_{0}}$ denoted as ${{\mathbf{A}}_{\Omega ,0}}$ (and ${{\mathbf{B}}_{\Omega ,0}}$), from the first $p$ columns from $\mathbf{A}$ (and $\mathbf{B}$) in $\mathsf{\mathcal{R}}{{\mathbf{x}}_{0}}\text{=}\mathbf{A\Lambda }{{\mathbf{B}}^{H}}$. Then, the VIP prior is further improved to ${{\mathbf{A}}_{\Omega ,l}}$ (and ${{\mathbf{B}}_{\Omega ,l}}$) with an updated solution ${{\mathbf{x}}_{l}}$ at the ${{l}^{th}}$ iteration in the implementation. From the modelling perspective, the proposed method seems to have the same DOF since this method still needs to determine the amplitude, phase, frequency, and damping factor from the measured data through Hankel SVD (HSVD). In practice, the proposed algorithm employs a two-loop update: The outer loop updates the VIP and the inner loop reconstructs the signal given the estimated VIP. Consequently, for the inner loop signal reconstruction, the updated VIP in the outer loop serves as the prior information, leading to a reduction of the nominal DOF. (The flowchart is presented in Supplement S2)

For the given ${{\mathbf{A}}_{\Omega,l}}$ and ${{\mathbf{B}}_{\Omega,l}}$, the model is solved with the Alternating Direction Method of Multipliers (ADMM) \cite{34}. The augmented Lagrange of Eq.(\ref{ex_6}) is
\begin{equation}
\begin{matrix}
\label{ex_7}
  G\left( \mathbf{x},\mathbf{Z},\mathbf{D} \right)={{\left\| \mathbf{Z} \right\|}_{*}}-\operatorname{Tr}\left( \mathbf{A}_{\Omega ,l}^{H}\mathbf{Z}{{\mathbf{B}}_{\Omega ,l}} \right)+ \\ 
  \langle \mathbf{D},\mathsf{\mathcal{R}}\mathbf{x}-\mathbf{Z}\rangle +\frac{\beta }{2}\left\| \mathsf{\mathcal{R}}\mathbf{x}-\mathbf{Z} \right\|_{F}^{2}+\frac{\lambda }{2}\left\| \mathbf{y}-\mathsf{\mathcal{U}}\mathbf{x}\right\|_{2}^{2}, \\ 
\end{matrix}
\end{equation}
where $\mathbf{D}$ is a dual variable, and $\left\langle \cdot ,\cdot  \right\rangle $ is the inner product in the Hilbert space of matrices, ${{\left\| \cdot  \right\|}_{F}}$ means the Frobenius norm and $\beta >0 $ is a parameter.

Eq.(\ref{ex_7}) are alternatingly solved via the following sub-problems until the algorithm converges:
\begin{equation}
\label{ex_8}
\left\{ \begin{aligned}
  & {{\mathbf{x}}^{k+1}}=\arg \underset{\mathbf{x}}{\mathop{\text{min}}}\,G\left( \mathbf{x},{{\mathbf{Z}}^{k}},{{\mathbf{D}}^{k}} \right) \\ 
 & {{\mathbf{Z}}^{k+1}}=\arg \underset{\mathbf{Z}}{\mathop{\text{min}}}\,G\left( {{\mathbf{x}}^{k+1}},\mathbf{Z},{{\mathbf{D}}^{k}} \right) \\ 
 & {{\mathbf{D}}^{k+1}}={{\mathbf{D}}^{k}}+\left( \mathsf{\mathcal{R}}{{\mathbf{x}}^{k+1}}-{{\mathbf{Z}}^{k+1}} \right) \\ 
\end{aligned} \right..  
\end{equation}

1) With fixed ${{\mathbf{D}}^{k}}$ and ${{\mathbf{Z}}^{k}}$, ${{\mathbf{x}}^{_{k+1}}}$ is obtained by solving
\begin{equation}
\label{deqn_ex9}
\underset{\mathbf{x}}{\mathop{\text{min}}}\,\left\langle {{\mathbf{D}}^{k}},\mathsf{\mathcal{R}}\mathbf{x}-{{\mathbf{Z}}^{k}} \right\rangle +\frac{\beta }{2}\left\| \mathsf{\mathcal{R}}\mathbf{x}-{{\mathbf{Z}}^{k}} \right\|_{F}^{2}+\frac{\lambda }{2}\left\| \mathbf{y}-\mathsf{\mathcal{U}}\mathbf{x} \right\|_{2}^{2},
\end{equation}
whose solution is 
\begin{equation}
\label{ex_10}
\begin{aligned}
\mathbf{x}^{k+1}= & \left(\lambda \mathcal{U}^* \mathcal{U}+\beta \mathcal{R}^* \mathcal{R}\right)^{-1} \\
& \left(\lambda \mathcal{U}^* \mathbf{y}+\beta \mathcal{R}^*\left(\mathbf{Z}^k-\frac{\mathbf{D}^k}{\beta}\right)\right)
\end{aligned}
\end{equation}
where the superscript $\text{*}$ denotes the adjoint operator.

2) With fixed ${{\mathbf{D}}^{k}}$ and ${{\mathbf{x}}^{k+1}}$ , ${{\mathbf{Z}}^{k+1}}$ is obtained by solving

\begin{equation}
\label{deqn_ex11}
\begin{aligned}
\min _{\mathbf{Z}}\|\mathbf{Z}\|_*-\operatorname{Tr}\left(\mathbf{A}_{\Omega, l}^H \mathbf{Z} \mathbf{B}_{\Omega, l}\right) & +\left\langle\mathbf{D}^k, \mathcal{R} \mathbf{x}^k-\mathbf{Z}\right\rangle \\
& +\frac{\beta}{2}\left\|\mathcal{R} \mathbf{x}^k-\mathbf{Z}\right\|_F^2,
\end{aligned}
\end{equation}

whose solution is 
\begin{equation}
\label{ex_12}
{{\mathbf{Z}}^{k+1}}\text{=}{{\mathcal{D}}_{1/\beta }}\left( \mathsf{\mathcal{R}}{{\mathbf{x}}^{k+1}}+\frac{1}{\beta }({{\mathbf{A}}_{\Omega ,l}}\mathbf{B}_{\Omega ,l}^{H}+{{\mathbf{D}}^{k}}) \right),
\end{equation}
where the ${{\mathcal{D}}_{1/\beta }}$ is a singular thresholding operator \cite{32} on a matrix with threshold $1/\beta $ .

3) By fixing ${{\mathbf{Z}}^{k+1}}$ and ${{\mathbf{x}}^{k+1}}$ , ${{\mathbf{D}}^{k+1}}$ is updated according to
\begin{equation}
\label{ex_13}
{{\mathbf{D}}^{k+1}}\text{=} {{\mathbf{D}}^{k}}+( \mathsf{\mathcal{R}}{{\mathbf{x}}^{k+1}}-{{\mathbf{Z}}^{k+1}}).
\end{equation}

The alternating in the three sub-equations of Eq.(\ref{ex_8}) stops if the number of iterations $l$ reaches the maximal number $L$, or ${{\xi }_{l+1}}=\left\| {{\mathbf{x}}_{l+1}}-{{\mathbf{x}}_{l}} \right\|/\left\| {{\mathbf{x}}_{l}} \right\|$, the normalized successive difference, is smaller than a given tolerance ${{\eta }_{\text{tol}}}\text{=}{{10}^{\text{-}5}}$.

The overall algorithm is summarized in Algorithm \ref{alg1}, including updating VIP subspaces ${{\mathbf{A}}_{\Omega }}$ and ${{\mathbf{B}}_{\Omega }}$ in the outer loop and solving the solution $\mathbf{x}$ in the inner loop.

\begin{algorithm}[tb]
	\renewcommand{\algorithmicrequire}{\textbf{Input:}}
	\renewcommand{\algorithmicensure}{\textbf{Output:}}
\caption{THE NUMERICAL ALGORITHM FOR THE PROPOSED METHOD}
\label{alg1}
\begin{algorithmic}[1]
\REQUIRE Initialization: $\mathbf{y}$, $\mathsf{\mathcal{R}}$, $\mathsf{\mathcal{U}}$, set outer maximal number of iterations \sethlcolor{green}\hl{}$L\text{=5}$ of updating as ${{\mathbf{A}}_{\Omega ,l}}$ and ${{\mathbf{B}}_{\Omega ,l}}$, convergence condition ${{\eta }_{\text{tol}}}={{10}^{-5}}$, and inner maximal number of iterations $K={{10}^{3}}$. Initialize the  solution ${{\mathbf{x}}_{0}}={{\mathsf{\mathcal{U}}}^{\text{*}}}\mathbf{y}$, the dual variable ${{\mathbf{D}}_{0}}=\mathbf{1}$, the number of iterations $k=0$, ${{\eta }^0}=1$, ${{\xi }_0}=1$\\
\textbf{While} (${{\xi }_{l}}\ge {{\eta }_{tol}}$) or ($l<L$), \textbf{do}
\begin{itemize}
\item[1)] 
Estimate ${{\mathbf{A}}_{\Omega ,l}}$ and ${{\mathbf{B}}_{\Omega ,l}}$ from the first $p$ columns of ${{\mathbf{A}}_{l}}$ and ${{\mathbf{B}}_{l}}$, respectively, where ${{\mathbf{A}}_{l}}$ and ${{\mathbf{B}}_{l}}$ obey the singular value decomposition $\mathsf{\mathcal{R}}{{\mathbf{x}}_{l}}={{\mathbf{A}}_{l}}\mathbf{\Lambda B}_{l}^{H}$. 

\textbf{While} (${{\eta }^{k}}\ge {{\eta }_{\text{tol}}}$) or ($k<K$), \textbf{do}
\begin{itemize}
    \item[a)] 
    Update ${{\mathbf{x}}^{k+1}}$   according to Eq.(\ref{ex_10});
    \item[b)] 
    Update ${{\mathbf{Z}}^{k+1}}$  according to Eq.(\ref{ex_12});
    \item[c)]
    Update ${{\mathbf{D}}^{k+1}}$  according to Eq.(\ref{ex_13});
    \item[d)]
    Compute ${{\eta }^{k+1}}=\left\| {{\mathbf{x}}^{k+1}}-{{\mathbf{x}}^{k}} \right\|/\left\| {{\mathbf{x}}^{k}} \right\|$  and  $k\leftarrow k+1$ ;
\end{itemize}
\textbf{End While} 
\item[2)] 
Set  $k=0$, ${{\eta }^{k}}=1$ and $l\leftarrow l+1$;
\item[3)] 
Update ${{\mathbf{x}}_{l+1}}\leftarrow {{\mathbf{x}}^{k+1}}$, ${{\xi }_{l+1}}=\left\| {{\mathbf{x}}_{l+1}}-{{\mathbf{x}}_{l}} \right\|/\left\| {{\mathbf{x}}_{l}} \right\|$;
\end{itemize}
\textbf{End While} 

\ENSURE The reconstructed FID $\mathbf{\hat{x}}\leftarrow {{\mathbf{x}}_{l+1}}$.\\

\end{algorithmic}
\end{algorithm}

\subsection{NMR Cloud Computing}
\begin{figure*}
\centering
\includegraphics[width=5.5in]{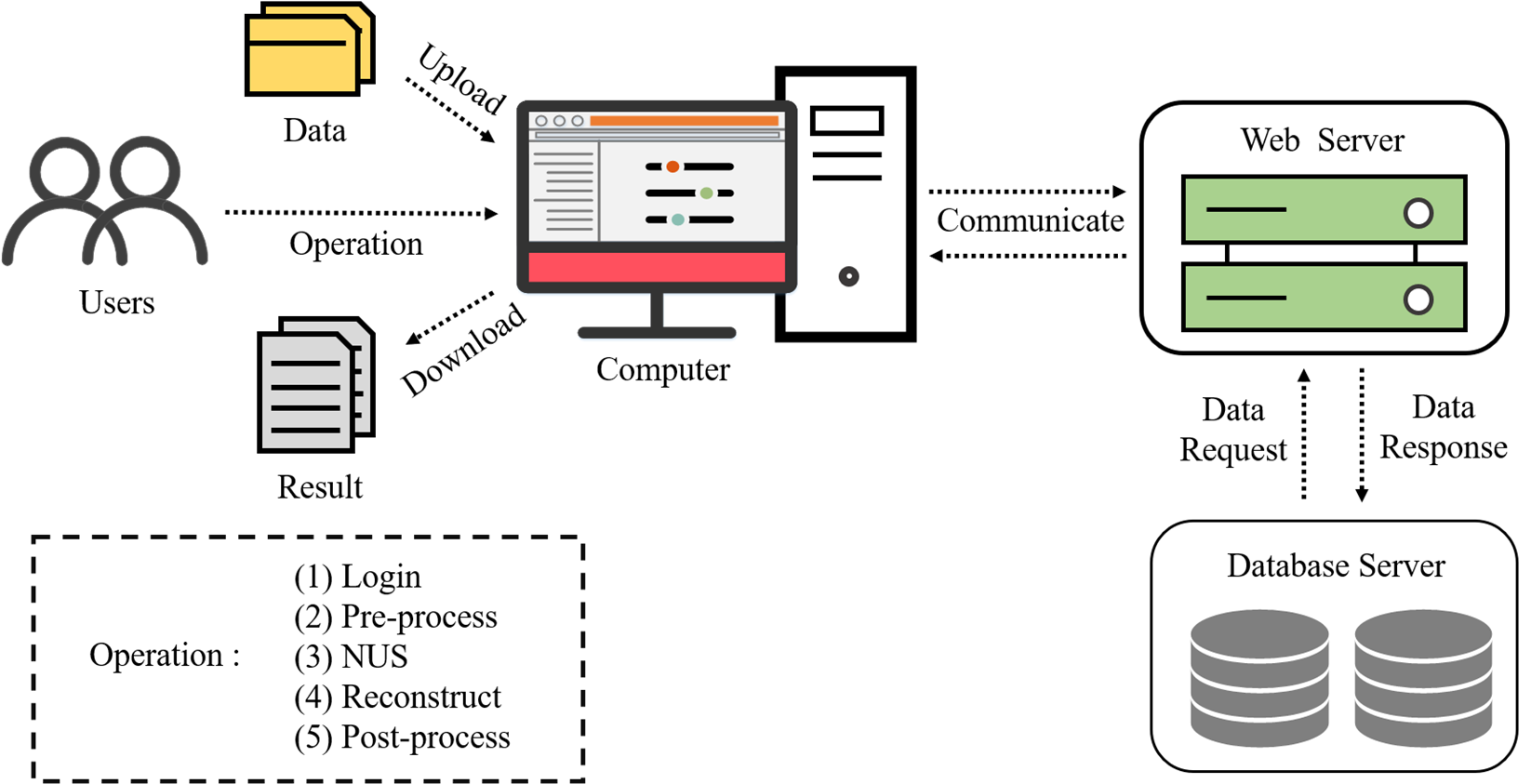}
\caption{A system framework of XCloud-NMR.}
\label{fig8}
\end{figure*}

Cloud computing is a state-of-the-art technology that is generally web-based and easily accessible at any time. Here, we develop XCloud-NMR, a cloud computing \cite{42} for spectra reconstruction with the proposed method. The XCloud-NMR employs a browser/service architecture (B/S), which comprises three main components (Figure \ref{fig8}): the browser, service, and data access layer (DAL). The browser component allows users to access the system directly through a web browser without any additional software installation. The interface design is user-friendly, with function buttons representing specific application programming interfaces (APIs) that call the next part of the service.
The service component uses Nginx in the background to transmit service requests to a distributed web server. This web server employs Google Remote Procedure Call (gRPC) to communicate and utilizes the Network File System (NFS) to share data.
The DAL component stores all data using an effective data storage strategy. MySQL is used to persistently store structured data such as usernames and passwords, while Redis is utilized to store temporary data that is frequently read and written.
Overall, the XCloud-NMR system's B/S architecture provides a convenient and efficient way for users to access the system and effectively handle service requests. The use of advanced technologies such as gRPC and NFS also ensures secure and reliable data communication and storage.

The whole processing flow is easy for NMR researchers: 1) Upload and pre-process raw data online; 2) Set the number of prior strong peaks and other reconstruction parameters; 3) Start online reconstruction; 4) Download the reconstructed data and show the spectrum. The manual, demo data, and post-processing scripts are accessible via the URL address and test accounts that are shared in Table \ref{tab3}.

\begin{table}[ht]
\begin{center}
\caption{TEST ACCOUNTS ON XCloud-NMR}
\label{tab3}
\begin{threeparttable}
\setlength{\tabcolsep}{10mm}{
\begin{tabular}{c c}
\toprule
User account &	Password \\
\midrule
CSG\_test001  & TEST@CSG01 \\
CSG\_test002  & USER\_test02 \\
CSG\_test003  & SERVICE\_CSGtest03 \\
\bottomrule
\end{tabular}}
  \begin{tablenotes}
        \footnotesize
        \item[*] The website of XCloud-VIP is \url{http://36.138.17.102:8989} %此处加入注释*信息
      \end{tablenotes}
  \end{threeparttable}
\end{center}
\end{table}
The spectra parameters and the reconstruction time are summarized in Table \ref{tab4}. The configuration of the local server includes two E5-2650v4 CPUs (12 cores) and 160 GB RAM. The cloud computing configuration is a CPU with 64 cores and 256 GB RAM. Table \ref{tab4} shows that the cloud computing enables shorter reconstruction time.
\begin{table*}[th!]
\begin{center}
\caption{IMPORTANT SPECTRAL PARAMETERS AND RECONSTRUCTION TIME}
\label{tab4}
\begin{threeparttable}
\begin{tabular}{ccccccc}
\toprule
Spectrum Type &	Sample & \makecell[c]{Spectra size for \\ reconstruction (${t}_{2}\times{t}_{1}$)}   &	References &	Sampling Rate &	\multicolumn{2}{c}{Reconstruction Time (s)} \\
\cline{6-7}	
\multicolumn{5}{c}{~} & ${{t}_{local}}$ ${{t}_{cloud}}$ (\textsl{AR}) \\
\midrule
HSQC &	GB1	 & ${1466}\times{170}$ &	Fig. \ref{fig5}	& 15\%  &	173.02 & 118.02 (31.8\%)\\
HSQC &	Ubiquitin &	${1024}\times{98}$ &	Fig. S3-1(a-1) &	20\% &	27.76 &	21.62 (22.1\%) \\
TROSY &	Ubiquitin &	${682}\times{128}$ &	Fig. S3-1(b-1) &	20\% &	47.79 &	33.54 (29.8\%) \\
NOESY &	Strychnine &	${2048}\times{256}$ &	Fig. \ref{fig7} &	20\% &	240.68 &	160.84 (33.2\%) \\
HSQC  &	  Mixture of the above last three chemicals & ${1024}\times{256}$ &	Fig. S5-1 &	15\% &	93.21 &	63.54 (31.8\%) \\
\bottomrule
\end{tabular}
  \begin{tablenotes}
        \footnotesize
        \item[*] The acceleration rate is defined as follow: 
        $ AR=\frac{{{t}_{local}}-{{t}_{cloud}}}{{{t}_{local}}}\times 100\%. $
  \end{tablenotes}
  \end{threeparttable}
\end{center}
\end{table*}

\section{RESULTS}
\subsection{Experimental Setup and Evaluation Criteria}
The performance of the proposed method is validated on realistic NMR data under the Poisson NUS pattern \cite{14}. The experimental details are summarized in Supplement S8. To avoid ambiguity caused by different field strength of different spectrometers, the parts per million (ppm) is defined as the unit of chemical shift \cite{35} according to:
\begin{equation}
\label{deqn_ex14}
\text{chemical shift(ppm) = }\frac{{{f}_{sample}}-{{f}_{ref}}}{{{f}_{spec}}}\times {{10}^{6}},
\end{equation}
where ${{f}_{sample}}$ is the resonance frequency of the sample, ${{f}_{ref}}$ is the absolute resonance frequency of a standard compound measured in the same magnetic field and ${{f}_{spec}}$ is the frequency of magnetic field strength of spectrometer.

To evaluate the quality of the reconstruction, we use the square of the Pearson correlation coefficient (${{\text{R}}^{2}}$) to measure the correlation between the reconstructed spectrum ${\mathbf{x}_{r}}$ and the fully sampled spectrum ${\mathbf{x}_{f}}$. A 2D spectrum is reshaped into a vector when the ${{\text{R}}^{2}}$ is measured. The ${{\text{R}}^{2}}$ is defined as follows:
\begin{equation}
\label{deqn_ex15}
{{\text{R}}^{2}}({{\mathbf{x}}_{r}},{{\mathbf{x}}_{f}})={{\left( \frac{cov({{\mathbf{x}}_{r}},{{\mathbf{x}}_{f}})}{{{\sigma }_{{{\mathbf{x}}_{r}}}}{{\sigma }_{{{\mathbf{x}}_{f}}}}} \right)}^{2}},	
\end{equation}
where $\operatorname{cov}\left( \cdot  \right)$ and ${{\sigma }_{{}}}$ denotes the covariance and standard deviation, respectively.

The relative $l_2$ norm error (RLNE) was used to quantitatively assess the accuracy of the signal reconstructions, according to

\begin{equation}
\begin{aligned}
& \operatorname{RLNE} \text { (ground-truth) }=\frac{\left\|\mathbf{x}_f-\mathbf{x}_r\right\|}{\left\|\mathbf{x}_f\right\|} \\
& \operatorname{RLNE} \text { (inner loop) }=\frac{\left\|\mathbf{x}^{k+1}-\mathbf{x}^k\right\|}{\left\|\mathbf{x}^k\right\|}, \\
& \operatorname{RLNE} \text { (outer loop) }=\frac{\left\|\mathbf{x}_{l+1}-\mathbf{x}_l\right\|}{\left\|\mathbf{x}_l\right\|}
\end{aligned}
\label{deqn_rlne}
\end{equation}
where $\|\cdot\|$ represents the $l_2$-norm. The $k$ and $l$ represents the number of inner and outer iterations for the proposed method, respectively. Unless explicitly stated otherwise, the term RLNE in this paper refers to RLNE (ground-truth).

The Signal Noise Ratio (SNR) is defined as
\begin{equation}
\text{SNR=10}\times \text{lg}\frac{\left\| {\mathbf{x}_{g}} \right\|_{2}^{2}}{\left\| {{\mathbf{x}}_{m}}-{{\mathbf{x}}_{g}} \right\|_{2}^{2}},
\end{equation}
where $\mathbf{x}_{g}$ and $\mathbf{x}_{m}$ are the ground truth data and measured data, respectively.
\subsection{Reconstruction of Simulation Data}
\subsubsection{Considerations for Real-world NMR Spectroscopy}
\begin{figure*}[!htb]
\centering
\includegraphics[width=6.7in]{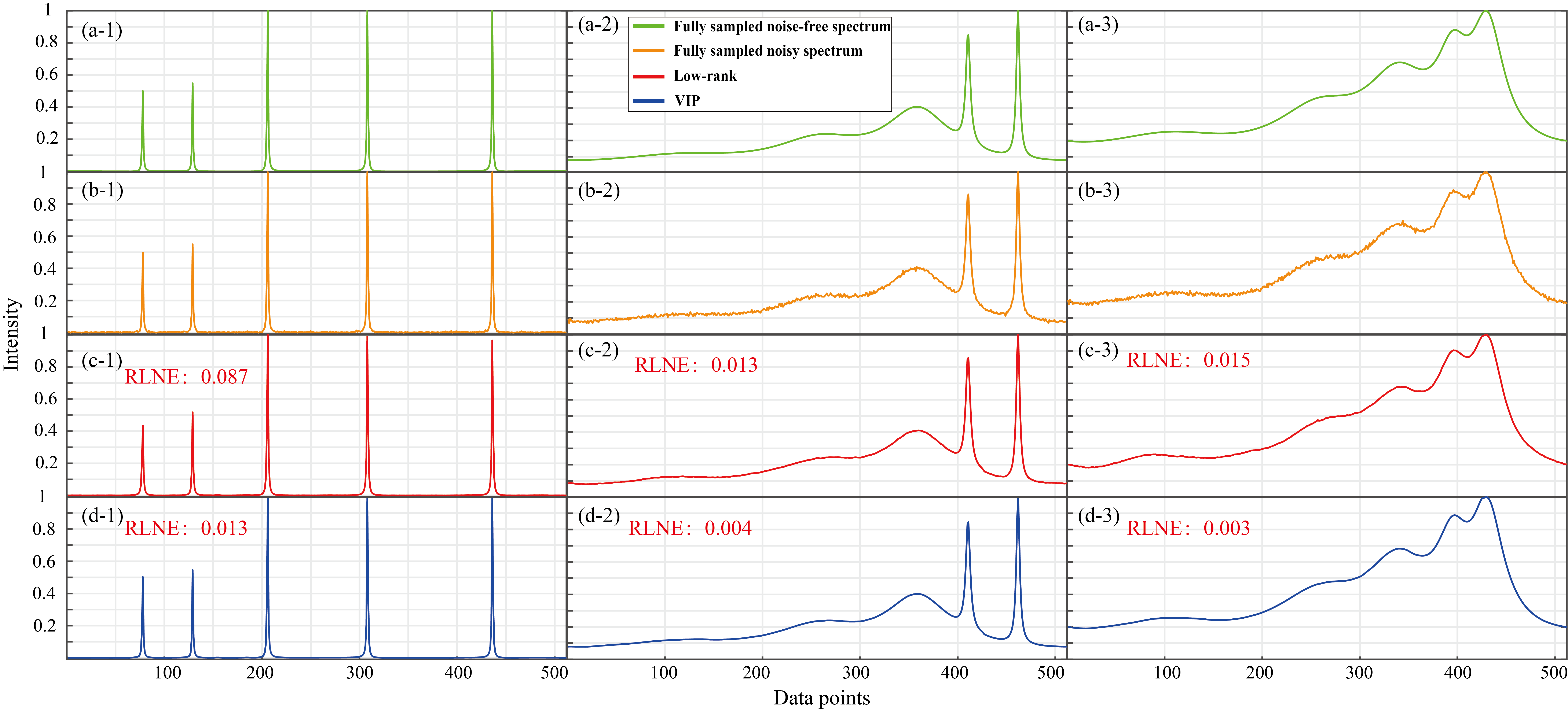}
\caption {The simulated data reconstruction with different scenarios when the acceleration factor (× 6.7) is moderate, i.e., the sampling rate is 15\%. (a-1)-(a-3), (b-1)-(b-3), (c-1)-(c-3) and (d-1)-(d-3) are the fully sampled noise-free spectra, the fully sampled noisy spectra, the reconstruction spectra by the low-rank method and the reconstruction by the proposed VIP method, respectively. Columns 1-3 are the spectra with non-equidistant resonances, different linewidths and line shapes, and overlapping signals, respectively. Note: The Gaussian noise with zero mean and a standard deviation of 0.005 is added to the FID.} 
\label{fig9a}
\end{figure*}

The first column of Figure \ref{fig9a} (Figures \ref{fig9a} (a-1)-(d-1)) present the results of non-equidistant spectra. Results show that the VIP holds the advantage over the low-rank method in preserving peaks. The $1^{st}$, $2^{nd}$, and the $5^{th}$ peaks of low-rank method have been weakened, while the proposed method preserves the same peak intensities as the fully sampled data. 

The second column of Figure \ref{fig9a} (Figures \ref{fig9a}  (a-2)-(d-2)) shows the results of different linewidth and line shapes of spectra. Both low-rank and VIP methods provide nice reconstruction. 

The last column of Figure \ref{fig9a} (Figures \ref{fig9a}  (a-3)-(d-3)) plot the reconstructions of the overlapping signals. Results show that the VIP has better performance, but the low-rank method performs sub-optimally on the reconstruction of heavily broadened and overlapped peaks (Figure \ref{fig9a}  (c-3)). The results indicate that the proposed method exhibits a lower reconstruction error (RLNE).

\subsubsection{Random Spectra}
The simulated NMR signal had different numbers of peaks, amplitude, phase, resonance frequency and $T_{2}^{*}$.
The simulated spectral parameters are listed in Table S9-5 (Supplement S9). We simulated five sets of spectra with varying number of peaks, ranging from 1 to 5, with each set containing 100 spectral signals. For each set where the number of spectral peaks was constant, the remaining parameters are randomly varied. Consequently, the total number of spectral signals is 500. Figure \ref{fig5a} shows the averaged performance of each set of spectra across 100 random NMR signals. The VIP achieves a more faithful reconstruction (lower reconstruction errors RLNEs and smaller variances) than the compared low-rank methods, indicating the good performance of the proposed method under different scenarios.
\begin{figure}[!t]
\centering
\includegraphics[width=3in]{Figure/Figure_5a.png}
\caption{Reconstruction errors of random spectra with 15\% of the fully sampled data. Note: Results from the low-rank method and the proposed VIP method are presented in red and blue, respectively. Note: The error bars are standard deviations of RLNEs with 100 signal trials. A smaller bar indicates a more robust reconstruction of random signals. The Gaussian noise with a mean of zero and a standard deviation of 0.005 is added to each set of data. }
\label{fig5a}
\end{figure}

\subsection{Reconstruction of Realistic NMR Data}
\begin{figure}[!t]
\centering
\includegraphics[width=3.5in]{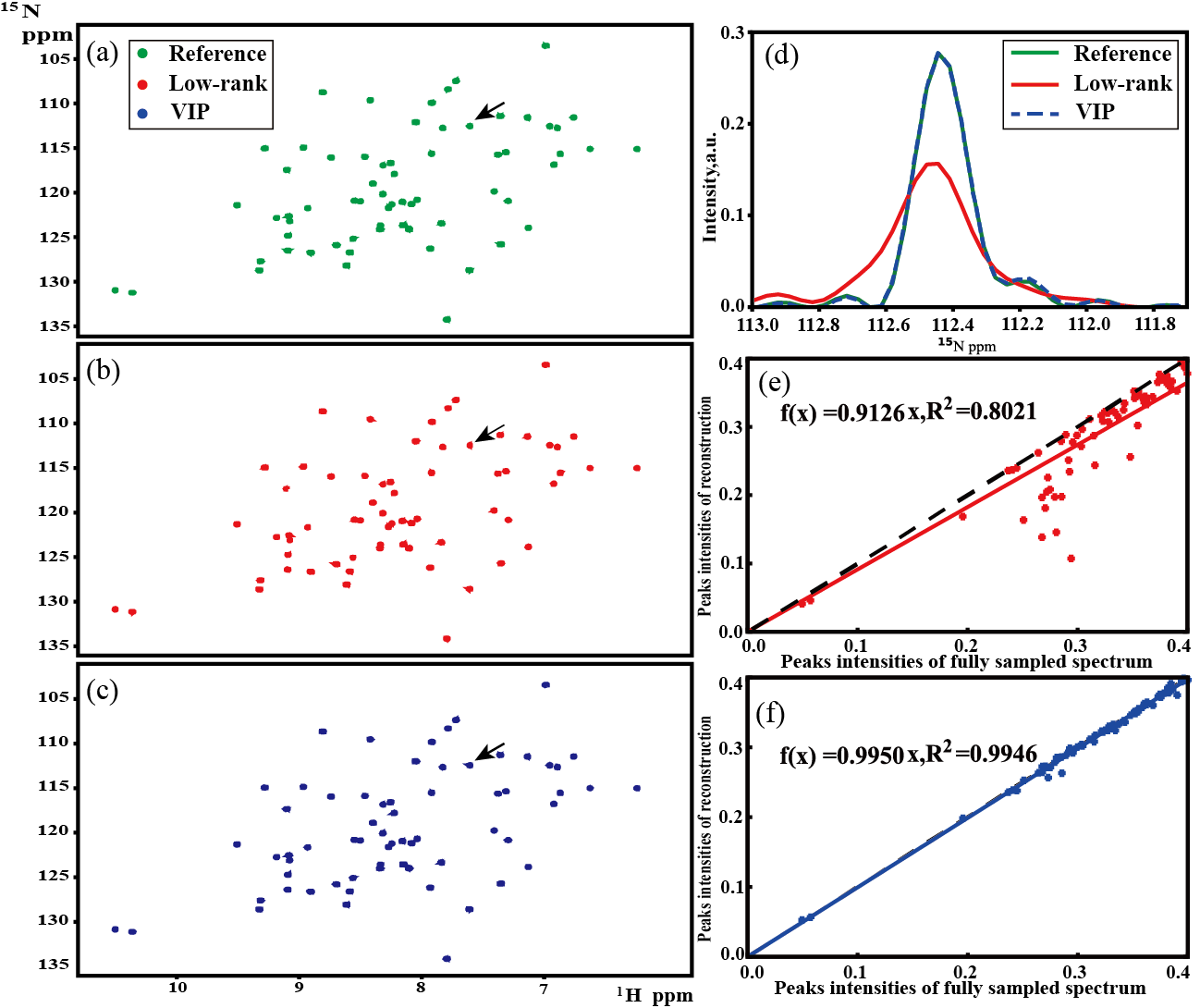}
\caption{Reconstruction of the 2D HSQC spectra of the protein GB1. (a) is the fully sampled reference spectrum, (b) and (c) are the reconstructed spectra from 15\% NUS data by the low-rank and the proposed VIP method, respectively, (d) is the zoomed-in 1D ${}^{\text{15}}\text{N}$ traces, and the green, red and blue lines represent the reference, low-rank and VIP reconstructed spectra, respectively. (e) and (f) are estimated with some peaks of low intensities at a range of [0, 0.4] using the low-rank and VIP, respectively. Note: The formula $f(x)=kx$ denotes the fitted curve according to peak intensities between the fully sampled data and the reconstructed data with a polynomial. The closer that the value of $k$ gets to 1, the smaller offset is between fully sampled data and reconstructed data is.}
\label{fig5}
\end{figure}

First NUS reconstruction is conducted on a 2D HSQC spectrum of a protein GB1 with a limited 15\% of fully sampled data. Figures \ref{fig5}(a)-(c) show that both the low-rank method and the proposed VIP can reconstruct most spectral peaks. However, a close look at the low-intensity peaks (Figures \ref{fig5}(d)) clearly shows that the low-rank method underestimates the intensities, while the VIP method achieves much higher fidelity peaks. Moreover, the low-rank method introduces a larger offset between the fully sampled data and the reconstruction data (Figures \ref{fig5}(e) and (f)). The correlation of low-intensity peaks is greatly increased from 0.80 in low-rank to 0.99 in VIP. Better reconstructions obtained with VIP are also observed on another two NMR spectra, including a ${}^{\text{1}}\text{H-}{}^{\text{15}}\text{N}$ best-TROSY spectrum and a ${}^{\text{1}}\text{H-}{}^{\text{15}}\text{N}$ HSQC spectrum of Ubiquitin (See Supplement S3).

\subsection{Quantitative Measures on Internuclear Distances}
\begin{figure}[!t]
\centering
\includegraphics[width=3.2in]{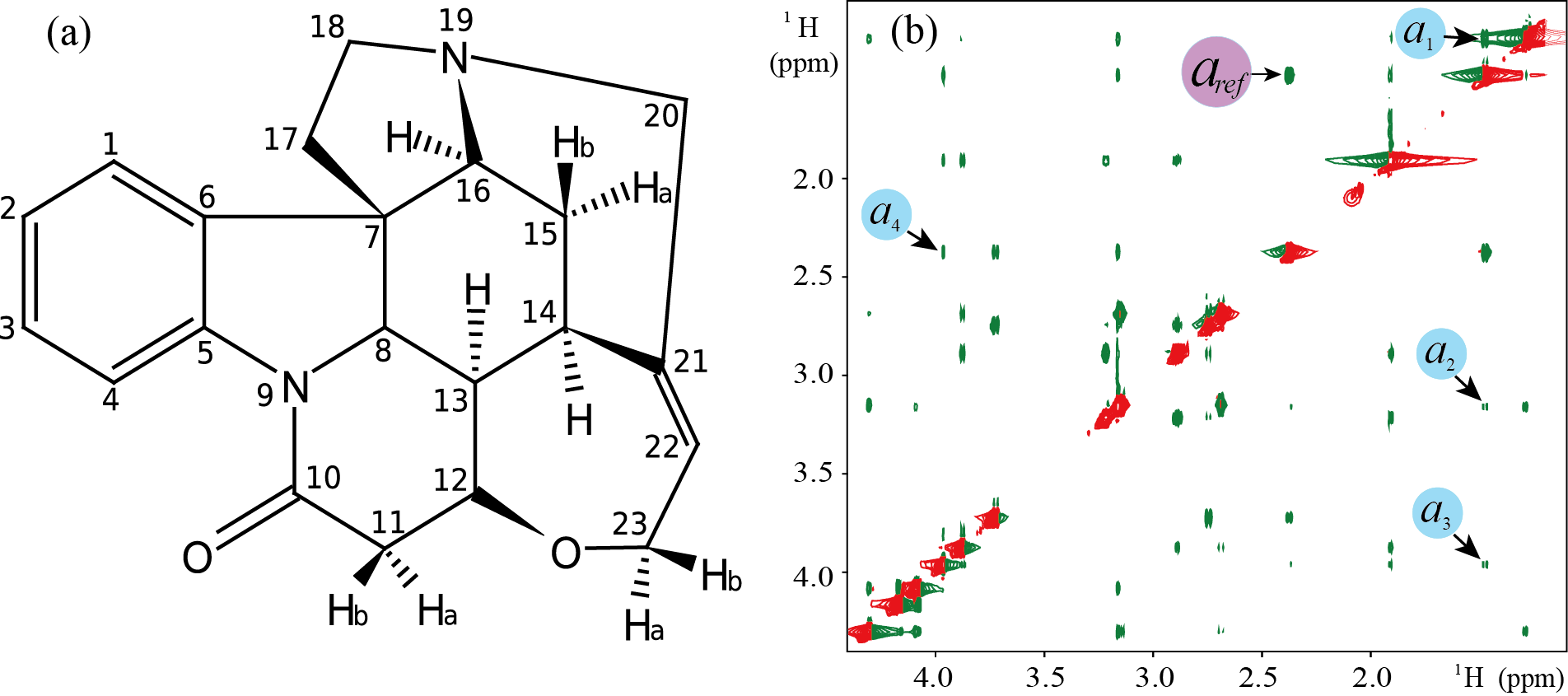}
\caption{Strychnine \cite{36,37}. (a) Molecular structure, (b) the fully sampled spectrum.}
\label{fig6}
\end{figure}

\begin{figure*}[!htb]
\centering
\includegraphics[width=6in]{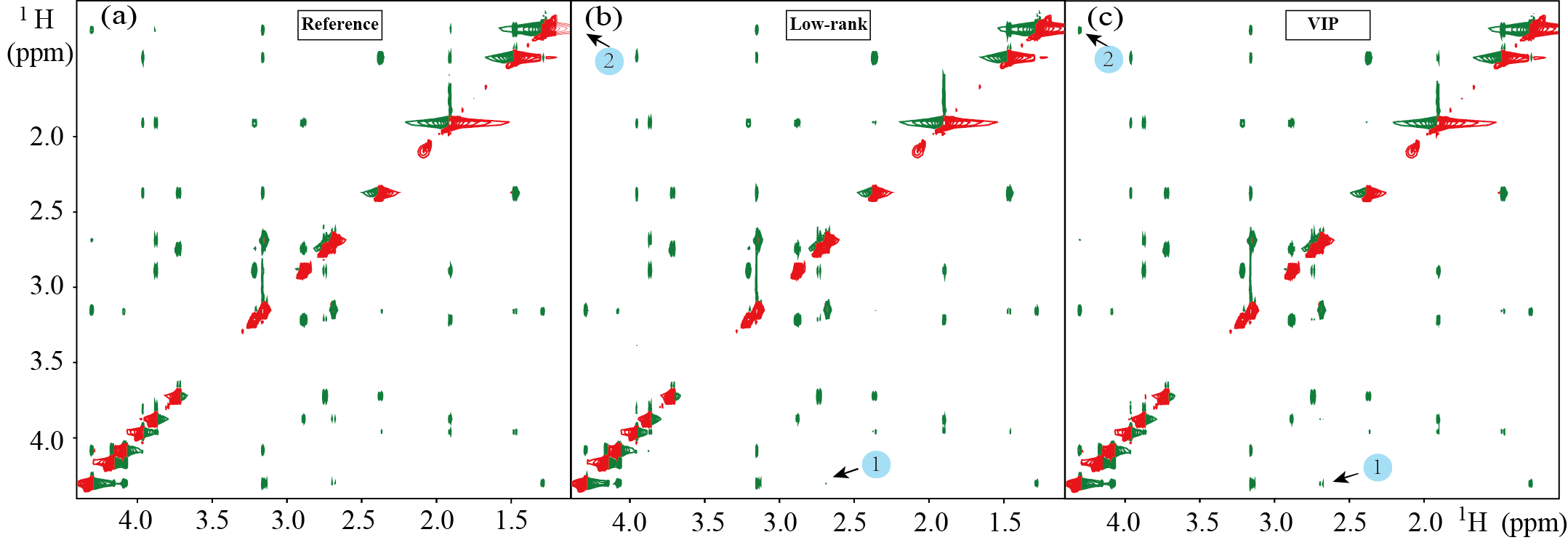}
\caption{Reconstruction of a 2D ${}^{\text{1}}\text{H-}{}^{\text{1}}\text{H}$ NOESY spectra of strychnine. (a) The fully sampled spectrum, (b) and (c) are reconstructed spectra by the low-rank and VIP, respectively. Note:  20\% of fully sampled data were used for the reconstructed spectra. Negative and positive peaks are represented with green and red colors, respectively.}
\label{fig7}
\end{figure*}

Quantitative measures on internuclear distances are analysed on a 2D ${}^{\text{1}}\text{H-}{}^{\text{1}}\text{H}$ NOESY spectrum of strychnine (Figure \ref{fig6}) \cite{36,37}. In the reconstruction from 20\% of the data with NUS, some cross-peaks (marked with arrows 1 and 2 in Figure \ref{fig7}(b)), which are missed or weakened by the low-rank method, are preserved well by the VIP method (Figure \ref{fig7}(c)). The correlation of peak intensity indicates that the VIP method can improve the fidelity of low-intensity peaks (Figure S4-1 of Supplement S4).
We further analyze the internuclear distance, which is important for computing molecular structure. The internuclear distance is defined as 
\begin{equation}
\label{deqn_ex16}
d={{d}_{ref}}{{\left(\frac{a}{{{a}_{ref}}} \right)}^{-\frac{1}{6}}},
\end{equation}
where ${d}_{ref}$=1.76 Å is a reference distance for internuclear H15a-H15b \cite{36,37}, ${a}_{ref}$ is the integral of the cross peak (marked as ${a}_{ref}$ in Figure \ref{fig6}(b)) that belongs to the internuclear H15a-H15b, and $a$ is the integral of the target cross-peaks (marked as ${a}_{1}$,${a}_{2}$,${a}_{3}$,${a}_{4}$ in Figure \ref{fig6}(b)). The 10-time Monte Carlo trial of NUS reconstruction was designed and the quantification results are summarized in Table \ref{tab1}. Both  the low-rank and the proposed method overestimate the internuclear distance. The reason is that undersampling reconstruction often leads to an underestimation of cross-peak intensities, which, in turn, results in an underestimation of peak integral values. Since peak integral values are inversely proportional to the internuclear distance, this leads to an overestimation of internuclear distances. However, the proposed method achieves much better and more stable estimation (closer to the ground-truth internuclear distance and with lower standard deviations of correlations).
\begin{table}[!t]
\begin{center}
\caption{SELECTED INTERNUCLEAR DISTANCES FROM THE NOESY SPECTRA OF STRYCHNINE (UNIT: Å)}
\label{tab1}
\begin{threeparttable}
\setlength{\tabcolsep}{0.05mm}{
\begin{tabular}{c c c c}
\toprule
Protons & ~~~	\makecell[c]{Fully sampled \\ spectrum}  & \multicolumn{2}{c}{Undersampling reconstruction} \\
\cline{3-4}	
\multicolumn{2}{c}{~} & Low-rank (Error) & ~~~VIP (Error) \\
\midrule
H15a-H15b &	1.76 &	1.76 (0.00\%) &	1.76 (0.00\%) \\
H13-H15a &	2.25 &	2.34 $\pm $ 0.09 (4.00\%)  & ~~~ 2.29 $\pm $ 0.03 (1.78\%) \\
H14-H15a &	2.68 &	2.80 $\pm $ 0.11 (4.48\%) & ~~~ 2.68 $\pm $ 0.04 (0.00\%) \\
H16-H15a &	2.54 &	2.72 $\pm $ 0.11 (7.09\%) & ~~~ 2.61 $\pm $ 0.05 (2.76\%) \\
H15b-H16 &	2.47 &	2.53 $\pm $ 0.04 (2.43\%) & ~~~ 2.50 $\pm $ 0.02 (1.21\%) \\
\bottomrule
\end{tabular}}

  \begin{tablenotes}
        \footnotesize
        \item[*]  The definition of Error is in Eq.(S5-2) (Supplement S5). For the numeric form A±S, A is the mean of correlation and S is the standard deviation of correlations over 10 Monte Carlo NUS trials. %此处加入注释*信息
      \end{tablenotes}
  \end{threeparttable}
\end{center}
\end{table}

\subsection{Quantitative Measures on the Relative Concentration}
Quantitative measures on the relative concentration are analysed on a mixture of 3 metabolites, including D-Glucose, $\beta$-Alanine and Valine (Figure S5-1). A series of ${\text{HSQC}_{i}}$ ($i=1,2,3$) spectra (Figure S5-2) are separately reconstructed by using 15\% of the data with NUS and then extrapolated back to a time-zero HSQC (${\text{HSQC}_{0}}$) spectrum. As the concentration of an individual metabolite is proportional to the peak intensity \cite{38}, the concentration measurement for an individual metabolite can be improved by averaging the intensities of multiple cross-peaks that belong to the same metabolite \cite{39}. A relative concentration of each metabolite is calculated as the ratio of its integration over the integration of the Valine (See Supplement S5 for more details).

Table \ref{tab2} indicates that VIP provides the closest estimation of concentration to that of the fully sampled spectrum.

\begin{table}[!ht]
\begin{center}
\caption{RELATIVE CONCENTRATIONS OF METABOLITES IN THE MIXTURE}
\label{tab2}
\begin{threeparttable}
\setlength{\tabcolsep}{1.5mm}{
\begin{tabular}{c c c c}
\toprule
Metabolites &	Fully sampled spectrum & \multicolumn{2}{c}{Undersampling reconstruction} \\
\cline{3-4}	
\multicolumn{2}{c}{~} & Low-rank (Error) & VIP (Error) \\
\midrule
Valine  &	1.00 &	1.00 (0.0\%) &	1.00 (0.0\%) \\
$\beta$-Alanine &	2.38 &	2.54 (6.7\%) &	2.40 (0.8\%) \\
D-Glucose &	4.60 &	4.84 (5.2\%) & 4.50 (2.2\%) \\
\bottomrule
\end{tabular}}

  \begin{tablenotes}
        \footnotesize
        \item[*] The concentration for each metabolite is improved by averaging the intensities of multiple, non-overlapping cross peaks assigned to that metabolite (Supplement S5). The closer the concentration to the fully sampling spectrum peak is, the more accurate the reconstruction method is.  The definition of Error is in Eq.(S5-2) (Supplement S5). %此处加入注释*信息
      \end{tablenotes}
  \end{threeparttable}
\end{center}
\end{table}

\subsection{Noise Level}
\begin{figure*}[t]
\centering
\includegraphics[width=6.5in]{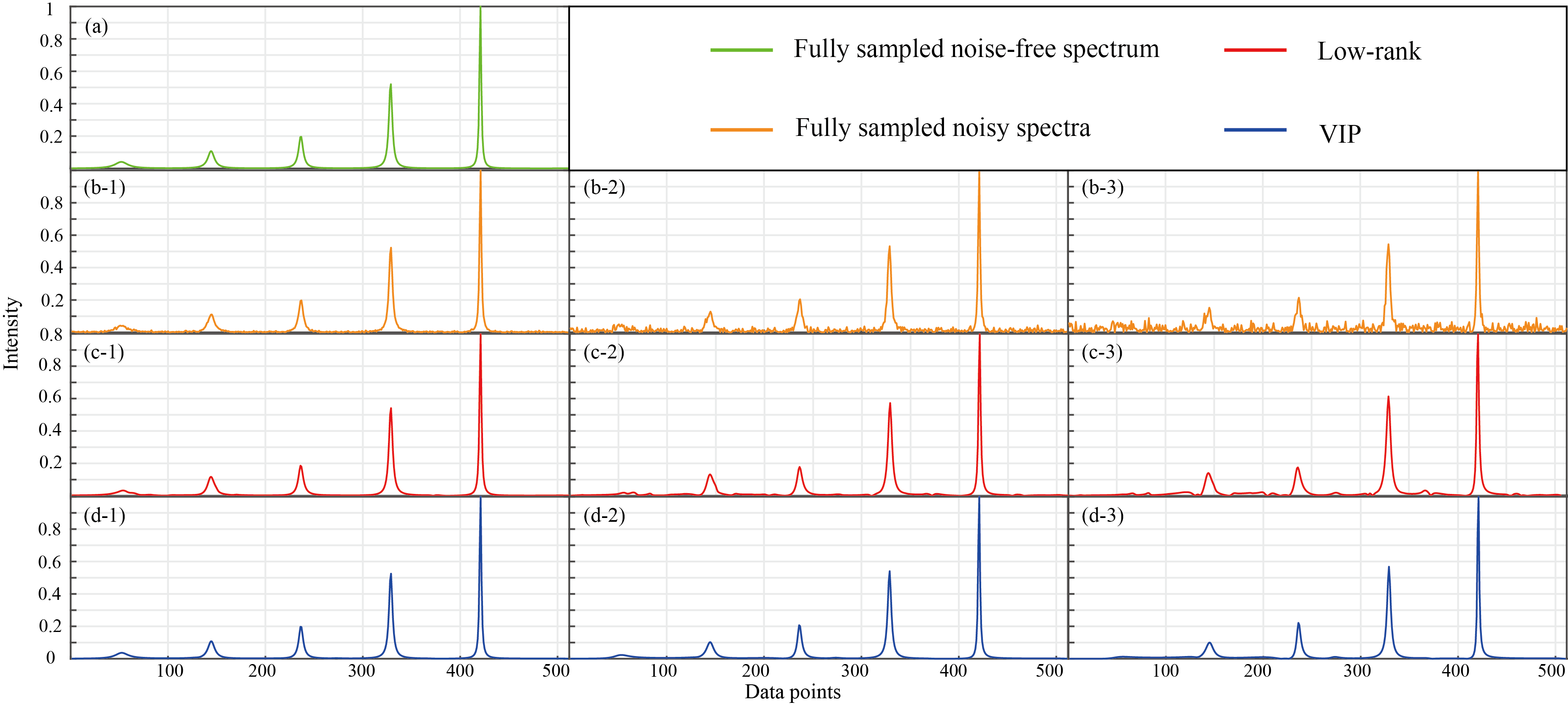}
\caption {Reconstructed spectra at different noise levels when acceleration factor (×10) is high, i.e., 10\% NUS rate. (a) is the same fully sampled noise-free spectrum. (b-1)-(b-3) are the fully sampled noisy spectra with noise standard deviations of 0.005, 0.02, and 0.04, respectively. (c-1)-(c-3) and (d-1)- (d-3) are the corresponding reconstructions obtained by low-rank and VIP, respectively. Note: The intensity of the lowest peak in the fully sampled noise-free spectrum is 0.04.}
\label{fig11a}
\end{figure*}

Spectra with different noise levels are added in Figure \ref{fig11a}. When the noise level is low, both low-rank and VIP methods produce high-quality spectra (Figures \ref{fig11a}(c-1) and (d-1)), even when the sampled data is severely limited, i.e., when the NUS rate is low (10\%). However, at a moderate or high noise level, the lowest intensity peak disappears in the low-rank (Figure \ref{fig11a} (c-2)) and VIP reconstruction method (Figure  \ref{fig11a} (d-2)). Then, increasing the NUS rate can lead to much better reconstruction (Figure S1-1 in Supplement S5), and the proposed method still outperforms the low-rank method on preserving the low-intensity peaks.

The SNR of different scenarios was shown in Table S1-1, it is obvious that the proposed method gains the highest SNR than the compared methods.

\subsection{Number of Prior Strong Peaks}
\begin{figure}[h]
\centering
\includegraphics[width=3.45in]{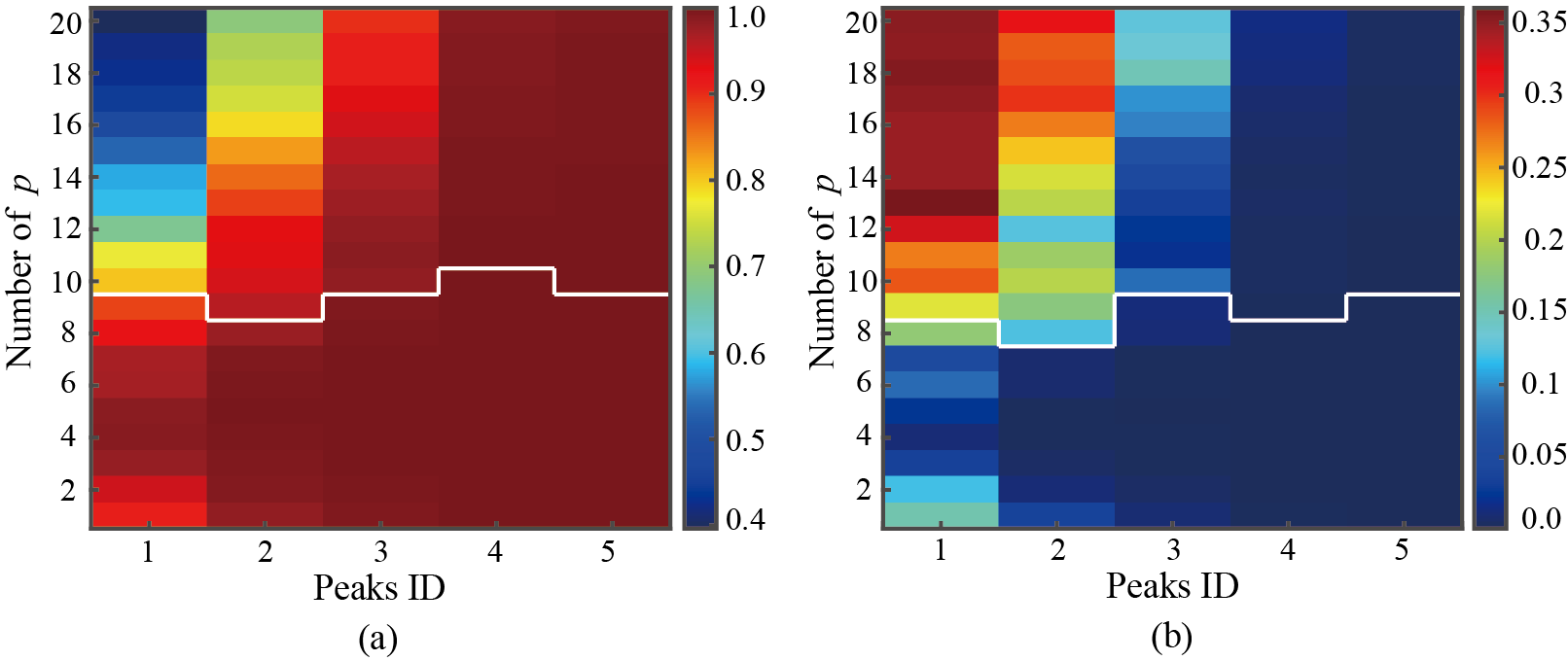}
\caption{The peak intensity correlation versus the number of strong peaks ($p$). (a) and (b) are the average and standard deviations for peak intensity correlation for each peak, respectively. The areas below the white line in (a) and (b) indicate a better performance (higher average or lower standard deviation) of VIP than that of the low-rank method. Note: The detailed correlations are reported in Table S6-1. The color bar in (a) and (b) indicates the peak intensity correlation, and standard deviation, respectively. The peaks 1$\sim$5 denote the peaks from the left to the right in Figure \ref{fig2}(a) of the main text. 8\% of the fully sampled data is used in the NUS. Note: The Gaussian noise with zero mean and a standard deviation of 0.005 are added to the FID.}
\label{fig10}
\end{figure}

The reconstruction performance of the proposed method under different numbers of strong peaks ($p$) is presented in Figure  \ref{fig10}. Figures \ref{fig10}(a) and (b) show that the VIP improves the reconstruction even when only one single strong peak is introduced. Best reconstruction performance is obtained when the number of strong peaks ($p$) is equal to the number of true peaks ($P$), i.e. $p=P=5$. The closer to the number of true peaks ($P$) is, the better the performance is.

Table S6-1 indicates that improved performance can always be obtained if the number of strong peaks ($p$) is set with in the range $p\in[1, 2P]$. In a practical application, when the number of peaks ($P$) is not known in advance, $p=1$ can still be set assuming that at least one peak ($p=1$) exists in the spectrum. With this setting, the proposed method can still obtain higher correlations of peak intensities than the compared method.

To sum up, VIP always outperforms the low-rank method if the number of strong peaks is between one and 2 times of the number of true peaks, indicating the robustness of this parameter.

\subsection{NUS Rate}
\begin{figure}[h]
    \centering
    \includegraphics[height=2.7in]{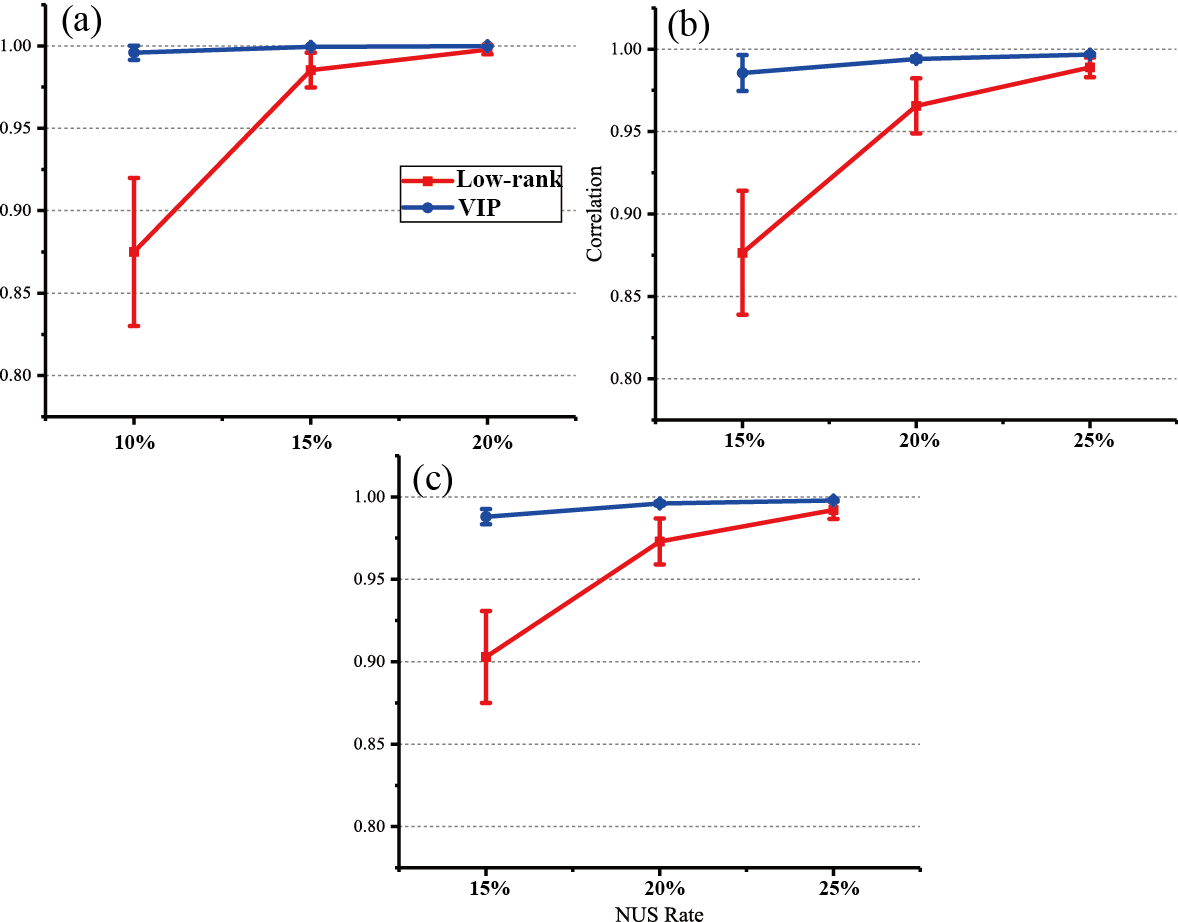}
    \caption{Peak intensity correlation for tested realistic spectra at different sampling rates. (a)-(c) are for the HSQC of GB1, HSQC of Ubiquitin and TROSY of Ubiquitin, respectively. Note: The error bars are the standard deviations of the correlations over 100 NUS sampling trials. The compressed sensing reconstructed spectrum is selected  as the initial reference.}
    \label{fig11}
\end{figure}

In this section, the reconstruction performance will be evaluated at different NUS rates. Results (Figure \ref{fig11}) indicate that the VIP significantly improves the correlation for all the tested data. Even with an extremely high acceleration factor (Figure \ref{fig11}(a)), e.g. the acceleration factor of 10 when the NUS rate is 10\%, the VIP increases the correlation to 0.99 from 0.87 obtained with low-rank. In addition, much lower standard deviations achieved by the VIP also indicates its high robustness to sampling rates.

\subsection{Automatic Reconstruction }
The automatic reconstruction include estimating noise level, automatic pre-reconstruction using low-rank approach under discrepancy principle (DP) \cite{51}, automatic estimation of strong peaks, and VIP reconstruction under DP.

First, noise level is automatically computed as the standard deviation of the tail of FID, assuming that noise-free FID signals decays nearly to zero at these data points. This is a common approach to automatic estimate the noise level \cite{44}. To verify that these FID data points conform to a Gaussian distribution, the Kolmogorov-Smirnov test is employed for validation. The probability was calculated with different truncated lengths under various noise standard deviations (SD) (SNRs: 24.8 $\pm$0.2dB (0.005), 12.8$\pm$0.2dB (0.02), 6.8$\pm$0.2dB (0.04)) and 100 Monte Carlo random experiments (Figure S7-1). The results indicate that the truncation point is approximately the last 20 points if the threshold for probability is set to 0.8. A relatively higher standard deviations of very small noise (19.7$\pm$0.2dB, (0.009$\pm$0.001)) is estimated for the ground-truth one (0.005). Other estimated standard deviations of noise are 0.024$\pm$0.004 (11.2$\pm$0.2dB) and 0.048$\pm$0.007 (5.2$\pm$0.2dB), which are close to the true ones 0.02 and 0.04, respectively.

Second, pre-reconstruction is performed on the undersampled FID using the conventional low-rank method \cite{13}. The regularization parameter $\lambda^*$ satisfies \cite{44}:
\begin{equation}
\left\|\mathbf{y}-\mathcal{U} \hat{\mathbf{x}}\left(\lambda^*\right)\right\|_2^2=C \sigma^2,
\label{deqn_ex21}
\end{equation}
where $\mathcal{U}$ is the undersampling operator, $\sigma$ is estimated noise standard deviation, $C$ is a constant that is equal to twice the number of acquired FID data points.

Third, the number of strong peaks is set to preserve the first $p$ largest singular values to maintain an energy loss according as
\begin{equation}
\frac{\left\|\mathbf{x}_{{rec}(p)}-\mathbf{x}_{{rec }}\right\|_2^2}{\left\|\mathbf{x}_{r e c}\right\|_2^2} \leq \delta,
\label{deqn_ex22}
\end{equation}
where $\mathbf{x}_{r e c(p)}$ and $\mathbf{x}_{r e c}$ represents the signal corresponding to the retained first $p$ largest singular values and all singular values in the pre-reconstruction, respectively \cite{13}. The energy loss is empirically set as a linear proportion of noise over the sampling rate according to $\delta=5 \sigma / S R$. This setting is reasonable in some sense because faithful strong peaks becomes fewer if noise increases or sampling rate decreases.

Last, the VIP reconstruction is done using DP, which means compute the regularization parameter ($\lambda^*$)  under DP first and then do VIP reconstruction with this regularization parameter. Both low-rank and VIP reconstruction done using DP is suboptimal since signal at the tail end of the FID does not conform exactly to a Gaussian noise.

\begin{figure*}[tb]
\centering
\includegraphics[width=5.8in]{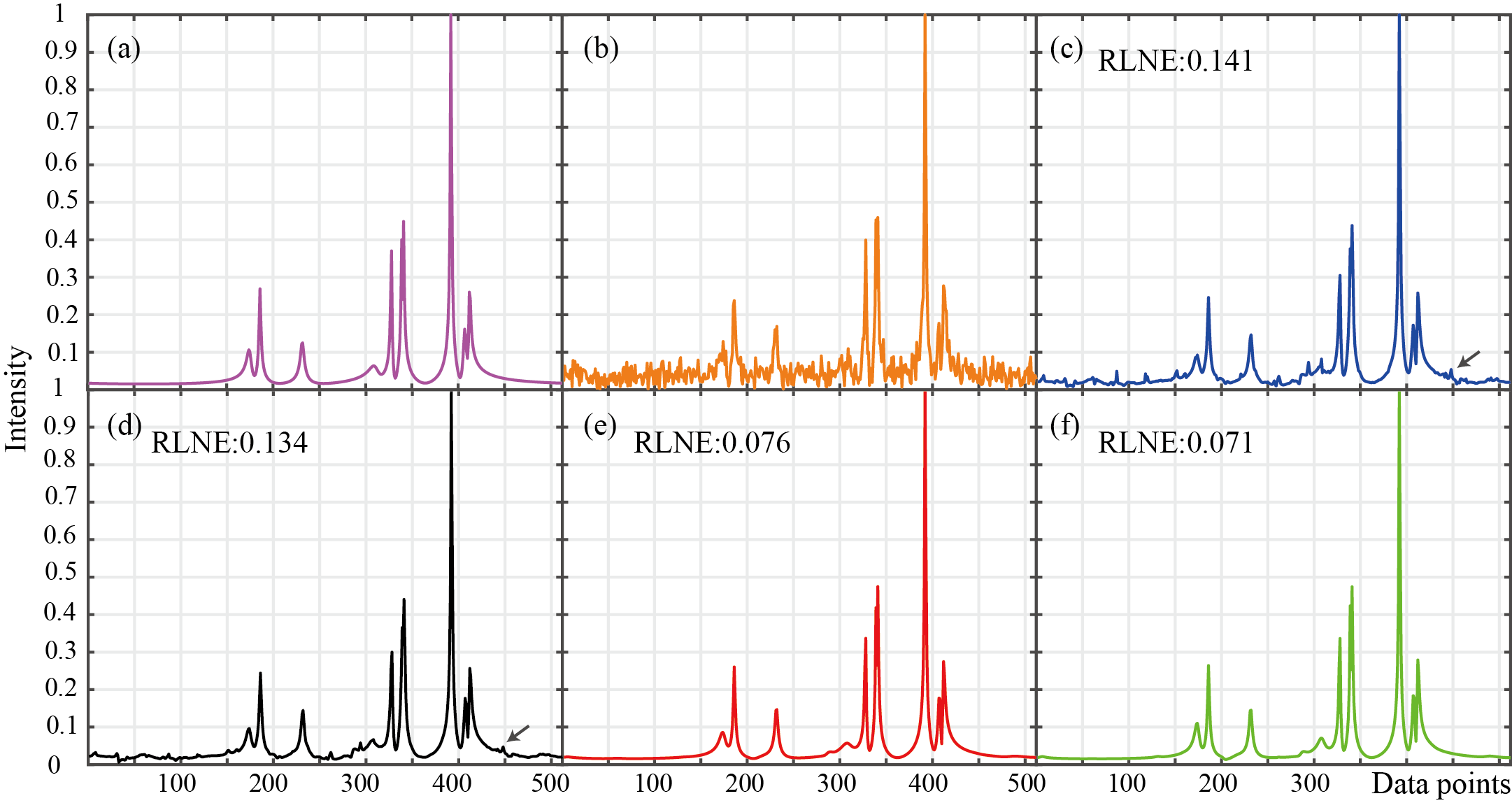}
\caption{Reconstruction of simulation spectrum with 10 peaks when the acceleration factor is 4 (i.e., the sampling rate is 25\%). (a) the noise-free fully sampled spectrum, (b) is the noisy fully sampled spectrum, (c) and (d) are the reconstructed spectra by the low-rank method with automatic parameter setting and optimal parameter settings, respectively, (e) and (f) are the reconstructed spectra by the proposed VIP method with automatic and optimal parameter settings, respectively. Note: The Gaussian noise with mean of zero and standard deviation of 0.02 is added to the FID. The optimal parameters are set reach the minimal reconstruction error, RLNE.}
\label{fig15}
\end{figure*}

Under different noise levels, we test the estimated number of strong peaks and reconstruction of synthetic spectra. The estimated number of peaks is always smaller than the true number and decreased if the noise level is increased (Table S7-1). This number approaches to the true one if the noise level is low and the sampling rate is high. Reconstructions of the 10-peak spectrum under three noise levels are shown in Figure S7-2 (low), Figure \ref{fig15} (moderate) and Figure S7-3 (high). Both low-rank and VIP methods reconstruct spectra very well if the noise level is low (Figure S7-2). When the noise level increases (Figure \ref{fig15}), small spurious peaks (Arrows in Figures \ref{fig15} (c) and (d)) are observed for the low-rank approach no matter under the automatic (The VIP reconstruction is done using DP without ground-truth signal.) or optimal parameter settings (The VIP reconstruction is done by minimizing the reconstruction error, RLNE, assuming the ground-truth signal is available). The proposed VIP improves reconstruction under the automatic parameter setting (Figure \ref{fig15}(e)) compared to the low-rank method under the optimal parameter setting, and shows a slight improvement under the optimal setting (Figure \ref{fig15}(f)). When heavy noise is introduced, automatic setting is sub-optimal for VIP (Figure S7-3) but the optimal setting still preserves the peaks much better.

\subsection{Empirical Convergence }
\begin{figure}[h]
\centering
\includegraphics[width=3.5in]{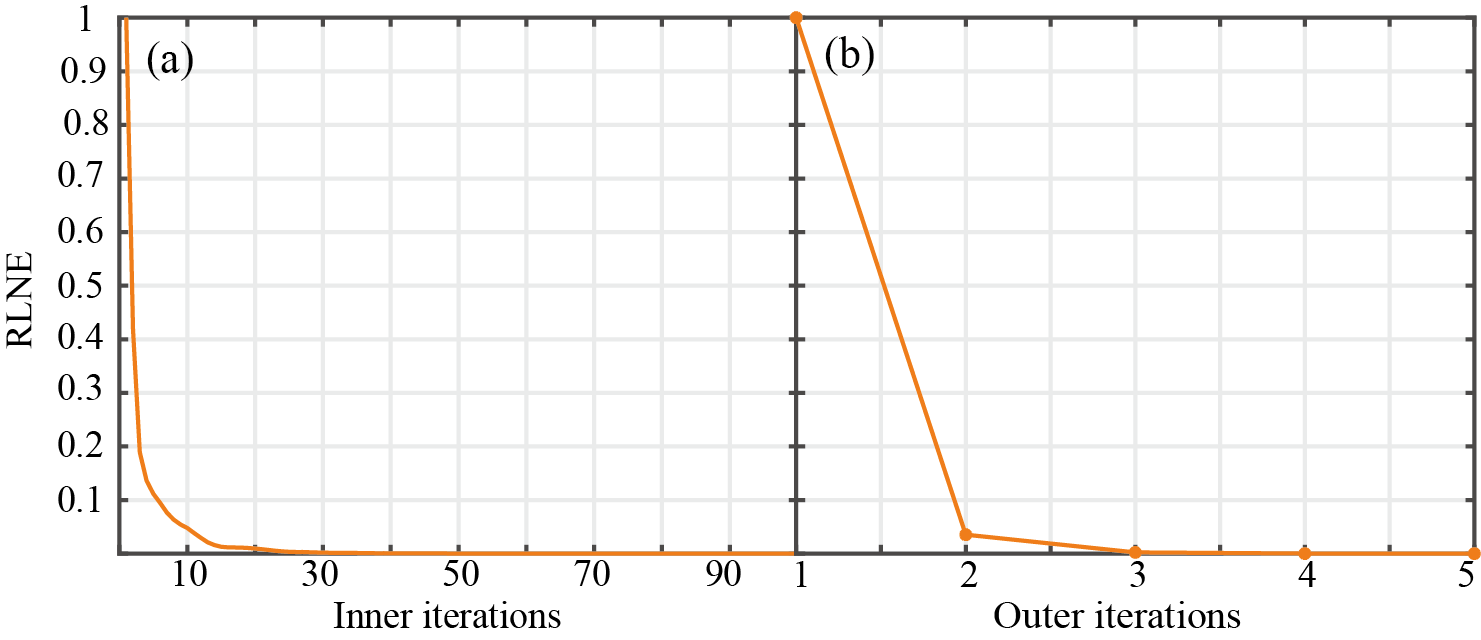}
\caption{RLNE changes with the increased number of iterations. (a) Inner iterations under the $1^{st}$ outer iteration, (b) all outer iterations. }
\label{fig17}
\end{figure}

The difference, quantified as RLNE (inner loop),  on the two reconstructions in the consecutive inner iterations, empirically converges (Figure \ref{fig17}(a)) and becomes smaller than $10^{-5}$ after 100 inner iterations. During the outer iterations, the RLNE (outer loop) reaches convergence criteria $10^{-5}$ at the $3^{rd}$ iteration and exhibits only slight improvement in the subsequent iterations (Figure \ref{fig17} (b)).

\section{DISCUSSION}
The discussion section primarily comprises three parts: 1. Data acquisition conditions, 2. Configuration of the reconstruction, and 3. Method's pros, cons, and limitations.

\subsection{Data Acquisition Conditions}
Data acquisition conditions include the setting of the NUS rate and a discussion on different noise levels.

NUS Rate: The acceleration factor achieved by VIP mainly depends on the physical properties of the specific spectrum. Higher spectral complexity requires reduced acceleration factor. To reconstruct high-fidelity realistic spectrum, whose peak intensity correlations are higher than 0.98 in Figure \ref{fig11}, the proposed method can reach an acceleration factor to 6$ \sim $10 but the low-rank method only allows an acceleration factor of 4$\sim $5.

Robustness to Noise: Noise is inevitable in the real world. For practical spectroscopy applications, meaningful peaks usually mean that the signal intensity is higher than the noise. If the noise intensity is higher than that of the meaningful peaks, it will be hard or even impossible to recover them when undersampling is applied. Figure \ref{fig11a} shows that the noise tolerance, i.e., the noise intensity, is lower than that of the meaningful peaks.

\subsection{Configuration of the Reconstruction}
An important aspect of the proposed method is how to configure the reconstruction, including the selection of number of prior strong peaks and initial reference spectra. 

Number of Strong Peaks: The number of strong peaks ($p$) is chosen empirically and the number of true peaks ($P$) is determined by the structure of tested biological samples. If no prior knowledge (structure) is provided, $p$ can be set to one assuming that at least one meaningful peak exists. The peak intensity correlations (Figure \ref{fig10}) obtained by the proposed VIP are always higher than those by the low-rank method, when the number of strong peaks ($p$) is smaller than two times of the number of true peaks ($P$).

\begin{figure*}[!htb]
\centering
\includegraphics[width=7in]{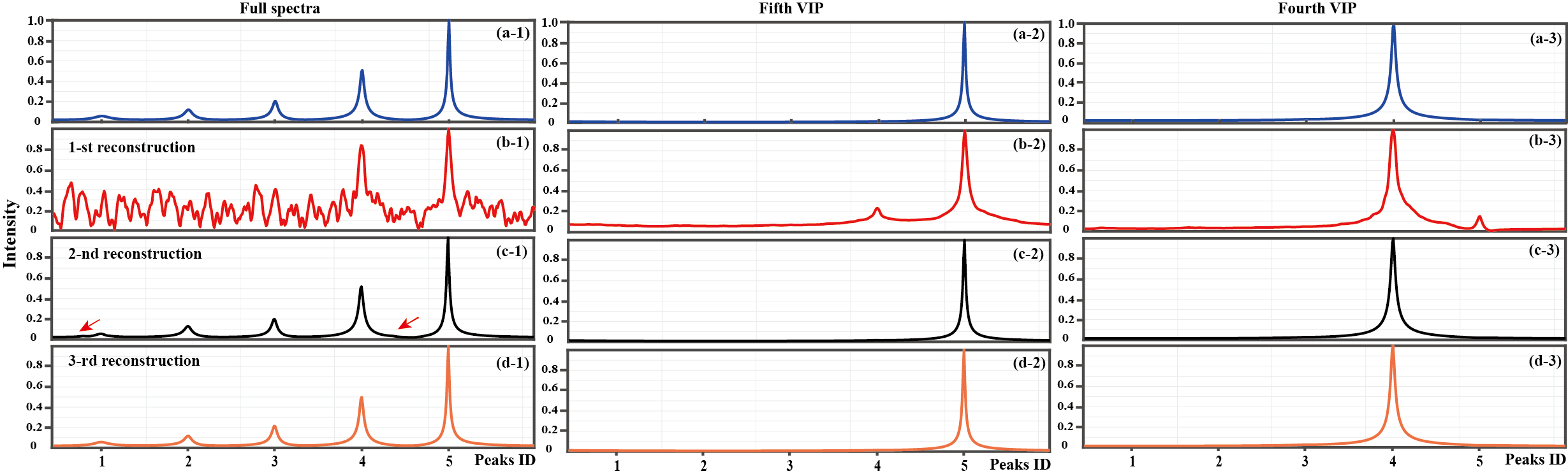}
\caption {Effect of updating VIP when the zero-filling spectrum is chosen as the initial reference. (a-1) the fully sampled spectrum; (b-1),(c-1) and (d-1) are the intermediate ${1}^{st}$, ${2}^{nd}$ and ${3}^{rd}$ reconstructions used to learn the VIP, respectively; (a-2) and (a-3) are accurate VIP learnt from the two strongest physical peaks in the fully sampled spectra in (a-1), respectively; (b-2) and (b-3) are estimated VIPs learnt from the two strongest peaks in the ${1}^{st}$ intermediate reconstruction in (b-1); (c-2) and (c-3) are estimated VIPs learnt from the two strongest peaks in the  ${2}^{nd}$ intermediate reconstruction in (c-1); (d-2) and (d-3) are the estimated VIPs learnt from the two strongest peaks in the ${3}^{rd}$ intermediate reconstruction in (d-1). Note: 8\% of fully sampled data were used in the reconstruction; All the reconstructions are accomplished by the VIP method with the number of strong peaks $p=3$; The Gaussian noise with zero mean and a standard deviation of 0.005 is added to the FID.}
\label{fig9}
\end{figure*}

Initial Reference Spectra: The estimated VIPs (Figures \ref{fig9}(b-2) and (b-3)) are learnt from the initial zero-filled spectrum (Figure \ref{fig9}(b-1)). These VIPs contain the prior information of spectral peaks, such as the resonance frequency and peak lineshape. This prior information is then introduced into the reconstruction model by VIPs and lead to high-quality reconstruction spectra (Figure \ref{fig9}(c-1)). However, these initial VIPs still have distortion, compared with the accurate VIPs (Figures \ref{fig9}(a-2) and (a-3)), and lead to distortions in the reconstruction spectra (marked with an arrow in Figure \ref{fig9}(c-1)). By further learning the VIP from the intermediate reconstruction (Figure \ref{fig9}(c-1)), the lineshapes of VIP are greatly improved (Figures \ref{fig9}(c-2) and (c-3)), and the high-quality spectrum (Figure \ref{fig9}(d-1)) is reconstructed. These observations imply that even starting from the zero-filled spectrum, the VIP becomes more reliable if more times of reference updating and VIP reconstructions. Our observations show that 5 times of learning is sufficient.

\subsection{Method's Pros, Cons, and Limitations}
This section primarily covers the advantages and disadvantages of the Hankel-based method in comparison to CS and DL, along with its limitations. Limitations include the reconstruction of non-Lorentzian lineshape and multiple resonance peaks.

The proposed approach exhibits distinct variations from other Hankel-based \cite{3,10,13,16,21,43}, CS \cite{8,31} and DLNMR \cite{11} methods in fast NUS NMR reconstruction. First, the proposed method preserves peak intensities (Figures \ref{fig9a} (a-1)-(d-1)) better than low-rank (Hankel-based) and CS methods because the proposed method combines both the singular value and subspace, while the other methods only focus on the singular value. Moreover, the benefits of proposed method and other Hankel-based method overcoming the fundamental limitation on the signal assumption in CS. The sparsity property (non-zero values in the spectrum) required by CS may not be satisfied well for challenging peaks (but the rank in Hankel method only depends on number of spectral peaks). Thus, peaks with different linewidths, lineshapes (Figures \ref{fig9a} (a-2)-(d-2)), and overlapping (Figures \ref{fig9a} (a-3)-(d-3)) can achieve robust reconstruction by the Hankel method. Besides, the proposed method having clearer interpretations than deep learning and not requiring prior database training. Thus, the Hankel method improves the sensitivity of practical fast sampling on NMR spectrometers. However, the limitations of proposed method is the long computation time owing to the big data with big matrix size and the system imperfections. These issues can be resolved through parallel computing \cite{19} and compensating for system deficiencies (More details are shown in S8).

\begin{figure*}[!htb]
\centering
\includegraphics[width=6.5in]{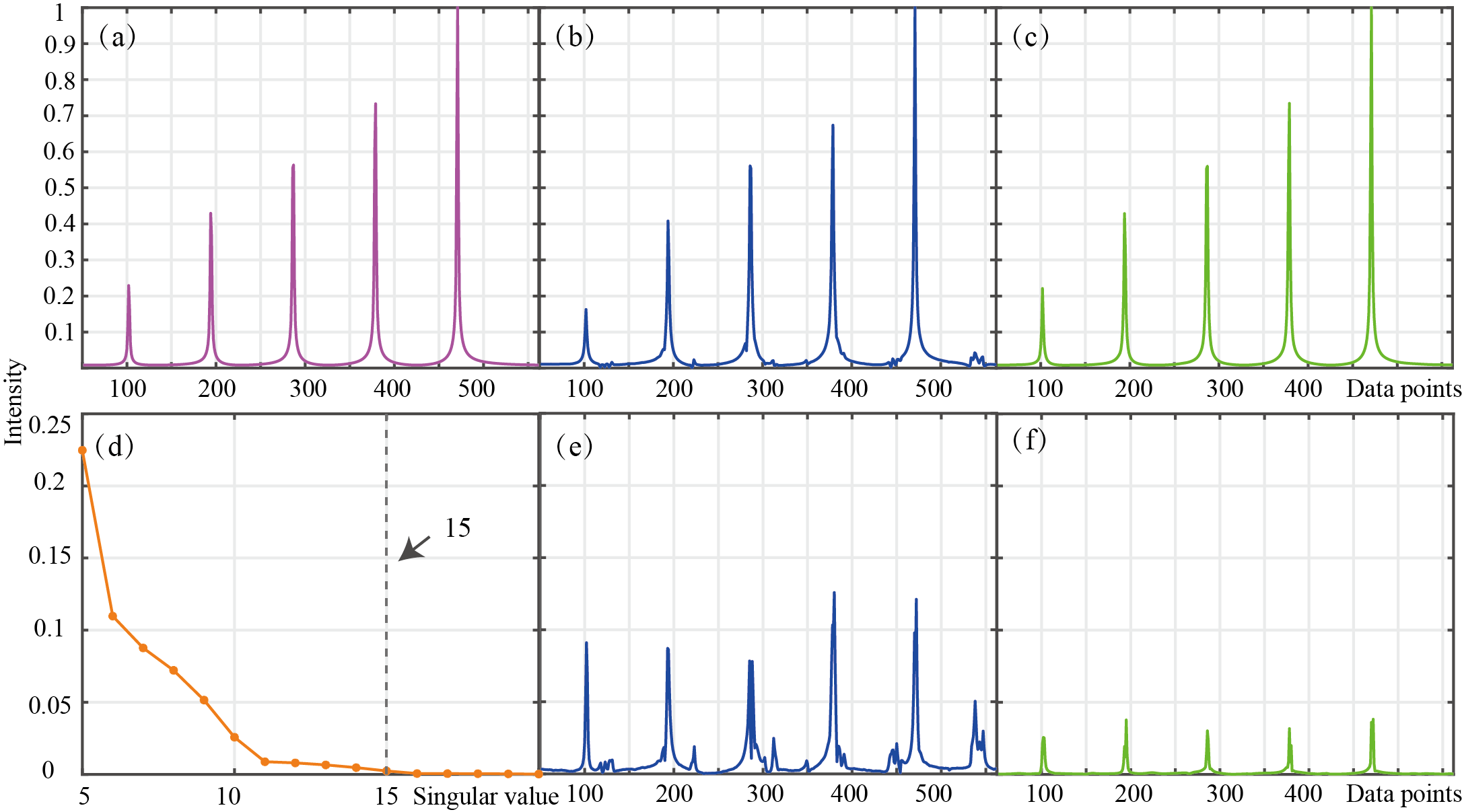}
\caption { Reconstructed Voigt-lineshape spectra when the acceleration factor is 12.5 (i.e., the sampling rate is 8\%). (a) the fully sampled noise-free spectrum, (b) and (c) are the reconstructed spectra by the low-rank method and the proposed VIP method, respectively. (d) is the singular value distribution corresponding to (a), (e) and (f) represent the reconstruction errors from the low-rank method and the proposed VIP method, respectively. Note: The Gaussian noise with mean of zero and standard deviation of 0.005 is added to the FID.}
\label{fig16}
\end{figure*}

Non-ideal Lorentzian Lineshape: The assumption of a Lorentzian lineshape for the resonance component has been widely used in fast NUS spectrum reconstruction \cite{10,13,16}. For NMR, especially the high-field NMR used in chemistry and biology, inhomogeneity and drift have been overcome much better for in vitro samples. To account for the presence of system imperfections,  we modeled the signal in Voigt-lineshape \cite{48}. A spectrum with five Voigt peaks  is simulated to test the reconstruction performance (Figure \ref{fig16}). Five peaks are denoted as 1 to 5 in Figure \ref{fig16} and their parameters are listed in Table S9-6. Figure \ref{fig16}(d) shows that the rank of this Hankel matrix ($R=15$) is larger than the number of peaks ($P=5$), due to the non-Lorentzian lineshape. In this case, neither the low-rank nor the proposed method successfully reconstruct spectra under a low sampling rate (8\%). Thus, this limitation of non-Lorentzian lineshape exists for both methods.
One way to mitigate the impact of non-Lorentzian lineshape is increasing the amount of sampled data. Figure S1-2 and Figure S1-3 show that reconstructed spectra become much better if the NUS rate is increased to 10\% and 25\%, respectively. When the undersampling rate is 10\%, the low-rank method exhibits spectral peak distortion (Figure S1-2(e)), while the proposed method yields better spectral peak than the low-rank method. However, there are still some errors (Figure S1-2(f)). As the sampling rate is further increased to 25\%, the proposed VIP method still yields smaller error (Figure 1-3(f)) than the low-rank method does (Figure S1-3(e)).
One possible solution may synthetize the mixed spectra that have Lorentzian, Gaussian and Voigt lineshapes and train a deep learning network \cite{11,41,49,50} that has high approximation ability to trained signals. For example, previous study has shown promises in reconstructing realistic spectrum that may have the non-Lorentzian lineshape, even with training on synthetized exponential signals (corresponding to the Lorentzian lineshape) \cite{11}.

Multiple Frequencies: Another limitation is reconstructing a spectrum with  multiple frequencies. Multiple frequencies mean that the rank of the Hankel matrix is relatively high, thus the low-rankness property, assumed in both the conventional low-rank or the proposed VIP methods, may not be satisfied well. A feasible solution is to increase the sampling rate (See detail in S1).

\section{CONCLUSION AND FUTURE WORK}
In this study, we propose an approach of self-adaptive virtual peak  to reconstruct high-quality NMR spectra with high acceleration factors, and set up an easy-accessible cloud computing platform, XCloud-NMR, for the proposed method. Virtual peaks incorporate the prior spectral information, such as the resonance frequency and peak lineshape into reconstructions. The proposed method can reconstruct reliable low-intensity peaks, and obtain faithful quantitative measures, such as internuclear distances and concentrations of mixtures. Thus, the proposed method enables higher acceleration factors of NMR data acquisition, which may significantly promote time-consuming NMR applications such as time-resolved experiments, real-time experiments, or \textit{in vivo} studies of short-lived systems. Future work could utilize the Vandermonde Hankel matrix decomposition  \cite{10}, which did not require the orthogonality of subspace but focus on the exponential function decomposition. The integrate physics simulation \cite{41} with deep learning, utilizing the power of deep learning to learn virtual peaks, would be another interesting try.

\section{Acknowledgments}
\noindent The authors are grateful to Qi Cao, and Hengfa Lu for plotting some figures, Vladislav Orekhov, Chunyan Xiong, Xinlin Zhang and Jinyu Wu for valuable discussions.

\footnotesize
\bibliographystyle{IEEEtran}
\bibliography{Manuscript_VIP,IEEEexample}

\begin{thebibliography}{10}
\providecommand{\url}[1]{#1}
\csname url@samestyle\endcsname
\providecommand{\newblock}{\relax}
\providecommand{\bibinfo}[2]{#2}
\providecommand{\BIBentrySTDinterwordspacing}{\spaceskip=0pt\relax}
\providecommand{\BIBentryALTinterwordstretchfactor}{4}
\providecommand{\BIBentryALTinterwordspacing}{\spaceskip=\fontdimen2\font plus
\BIBentryALTinterwordstretchfactor\fontdimen3\font minus \fontdimen4\font\relax}
\providecommand{\BIBforeignlanguage}[2]{{%
\expandafter\ifx\csname l@#1\endcsname\relax
\typeout{** WARNING: IEEEtran.bst: No hyphenation pattern has been}%
\typeout{** loaded for the language `#1'. Using the pattern for}%
\typeout{** the default language instead.}%
\else
\language=\csname l@#1\endcsname
\fi
#2}}
\providecommand{\BIBdecl}{\relax}
\BIBdecl

\bibitem{1}
M.~Mobli and J.~C. Hoch, ``Nonuniform sampling and non-fourier signal processing methods in multidimensional {NMR},'' \emph{Prog. Nucl. Magn. Reson. Spectrosc.}, vol.~83, pp. 21--41, 2014.

\bibitem{2}
S.~Robson, H.~Arthanari, S.~G. Hyberts, and G.~Wagner, ``Nonuniform sampling for {NMR} spectroscopy,'' \emph{Methods Enzymol.}, vol. 614, pp. 263--291, 2019.

\bibitem{3}
T.~Qiu, Z.~Wang, H.~Liu, D.~Guo, and X.~Qu, ``Review and prospect: {NMR} spectroscopy denoising and reconstruction with low-rank {Hankel} matrices and tensors,'' \emph{Magn. Reson. Chem.}, vol.~59, no.~3, pp. 324--345, 2021.

\bibitem{4}
B.~E. Coggins, R.~A. Venters, and P.~Zhou, ``Radial sampling for fast {NMR}: concepts and practices over three decades,'' \emph{Prog. Nucl. Magn. Reson. Spectrosc.}, vol.~57, no.~4, p. 381, 2010.

\bibitem{5}
X.~Qu, X.~Cao, D.~Guo, and Z.~Chen, ``Compressed sensing for sparse magnetic resonance spectroscopy,'' vol.~10, p. 3371, 2010.

\bibitem{6}
X.~Qu, D.~Guo, X.~Cao, S.~Cai, and Z.~Chen, ``Reconstruction of self-sparse 2{D} {NMR} spectra from undersampled data in the indirect dimension,'' \emph{Sensors}, vol.~11, no.~9, pp. 8888--8909, 2011.

\bibitem{7}
D.~J. Holland, M.~J. Bostock, L.~F. Gladden, and D.~Nietlispach, ``Fast multidimensional {NMR} spectroscopy using compressed sensing,'' \emph{Angew. Chem. Int. Ed.}, vol. 123, no.~29, pp. 6678--6681, 2011.

\bibitem{8}
K.~Kazimierczuk and V.~Y. Orekhov, ``Accelerated {NMR} spectroscopy by using compressed sensing,'' \emph{Angew. Chem. Int. Ed.}, vol.~50, no.~24, pp. 5556--5559, 2011.

\bibitem{9}
J.~Ying, F.~Delaglio, D.~A. Torchia, and A.~Bax, ``Sparse multidimensional iterative lineshape-enhanced (smile) reconstruction of both non-uniformly sampled and conventional {NMR} data,'' \emph{J. Biomol. NMR}, vol.~68, no.~2, pp. 101--118, 2017.

\bibitem{10}
J.~Ying, J.-F. Cai, D.~Guo, G.~Tang, Z.~Chen, and X.~Qu, ``Vandermonde factorization of {Hankel} matrix for complex exponential signal recovery—application in fast {NMR} spectroscopy,'' \emph{IEEE Trans. Signal Process.}, vol.~66, no.~21, pp. 5520--5533, 2018.

\bibitem{11}
X.~Qu, Y.~Huang, H.~Lu, T.~Qiu, D.~Guo, T.~Agback, V.~Orekhov, and Z.~Chen, ``Accelerated nuclear magnetic resonance spectroscopy with deep learning,'' \emph{Angew. Chem. Int. Ed.}, vol. 132, no.~26, pp. 10\,383--10\,386, 2020.

\bibitem{12}
D.~Chen, Z.~Wang, D.~Guo, V.~Orekhov, and X.~Qu, ``Review and prospect: Deep learning in nuclear magnetic resonance spectroscopy,'' \emph{Chem.--Eur. J.}, vol.~26, no.~46, pp. 10\,391--10\,401, 2020.

\bibitem{13}
X.~Qu, M.~Mayzel, J.-F. Cai, Z.~Chen, and V.~Orekhov, ``Accelerated {NMR} spectroscopy with low-rank reconstruction,'' \emph{Angew. Chem. Int. Ed.}, vol.~54, no.~3, pp. 852--854, 2015.

\bibitem{14}
S.~G. Hyberts, A.~G. Milbradt, A.~B. Wagner, H.~Arthanari, and G.~Wagner, ``Application of iterative soft thresholding for fast reconstruction of {NMR} data non-uniformly sampled with multidimensional poisson gap scheduling,'' \emph{J. Biomol. NMR}, vol.~52, no.~4, pp. 315--327, 2012.

\bibitem{15}
X.~Qu, T.~Qiu, D.~Guo, H.~Lu, J.~Ying, M.~Shen, B.~Hu, V.~Orekhov, and Z.~Chen, ``High-fidelity spectroscopy reconstruction in accelerated {NMR},'' \emph{Chem. Commun.}, vol.~54, no.~78, pp. 10\,958--10\,961, 2018.

\bibitem{16}
J.~Ying, H.~Lu, Q.~Wei, J.-F. Cai, D.~Guo, J.~Wu, Z.~Chen, and X.~Qu, ``Hankel matrix nuclear norm regularized tensor completion for $ n $-dimensional exponential signals,'' \emph{IEEE Trans. Signal Process.}, vol.~65, no.~14, pp. 3702--3717, 2017.

\bibitem{17}
J.~C. Hoch, ``{NMR} {Data} {Processing},'' \emph{Phys. Med. Biol.}, p. 611, 1997.

\bibitem{18}
H.~M. Nguyen, X.~Peng, M.~N. Do, and Z.-P. Liang, ``Denoising {MR} spectroscopic imaging data with low-rank approximations,'' \emph{IEEE Trans. Biomed. Eng.}, vol.~60, no.~1, pp. 78--89, 2012.

\bibitem{19}
D.~Guo, H.~Lu, and X.~Qu, ``A fast low rank {Hankel} matrix factorization reconstruction method for non-uniformly sampled magnetic resonance spectroscopy,'' \emph{IEEE Access}, vol.~5, pp. 16\,033--16\,039, 2017.

\bibitem{20}
D.~Guo and X.~Qu, ``Improved reconstruction of low intensity magnetic resonance spectroscopy with weighted low rank {Hankel} matrix completion,'' \emph{IEEE Access}, vol.~6, pp. 4933--4940, 2018.

\bibitem{21}
H.~Lu, X.~Zhang, T.~Qiu, J.~Yang, J.~Ying, D.~Guo, Z.~Chen, and X.~Qu, ``Low rank enhanced matrix recovery of hybrid time and frequency data in fast magnetic resonance spectroscopy,'' \emph{IEEE Trans. Biomed. Eng.}, vol.~65, no.~4, pp. 809--820, 2017.

\bibitem{22}
P.~Koehl, ``Linear prediction spectral analysis of {NMR} data,'' \emph{Prog. Nucl. Magn. Reson. Spectrosc.}, vol.~34, no. 3-4, pp. 257--299, 1999.

\bibitem{23}
J.~P. Haldar, ``Low-rank modeling of local $ k $-space neighborhoods {(LORAKS)} for constrained {MRI},'' \emph{IEEE Trans. Med. Imaging}, vol.~33, no.~3, pp. 668--681, 2013.

\bibitem{24}
K.~H. Jin, D.~Lee, and J.~C. Ye, ``A general framework for compressed sensing and parallel {MRI} using annihilating filter based low-rank {Hankel} matrix,'' \emph{IEEE Trans. Comput. Imaging}, vol.~2, no.~4, pp. 480--495, 2016.

\bibitem{25}
G.~Ongie and M.~Jacob, ``A fast algorithm for convolutional structured low-rank matrix recovery,'' \emph{IEEE Trans. Comput. Imaging}, vol.~3, no.~4, pp. 535--550, 2017.

\bibitem{26}
X.~Zhang, D.~Guo, Y.~Huang, Y.~Chen, L.~Wang, F.~Huang, Q.~Xu, and X.~Qu, ``Image reconstruction with low-rankness and self-consistency of $k$-space data in parallel {MRI},'' \emph{Med. Image Anal.}, vol.~63, p. 101687, 2020.

\bibitem{27}
F.~Lam, Y.~Li, and X.~Peng, ``Constrained magnetic resonance spectroscopic imaging by learning nonlinear low-dimensional models,'' \emph{IEEE Trans. Med. Imaging}, vol.~39, no.~3, pp. 545--555, 2019.

\bibitem{28}
Y.~Li, Y.~Zhao, R.~Guo, T.~Wang, Y.~Zhang, M.~Chrostek, W.~C. Low, X.-H. Zhu, Z.-P. Liang, and W.~Chen, ``Machine learning-enabled high-resolution dynamic deuterium {MR} spectroscopic imaging,'' \emph{IEEE Trans. Med. Imaging}, vol.~40, no.~12, pp. 3879--3890, 2021.

\bibitem{40}
X.~Zhang, H.~Lu, D.~Guo, Z.~Lai, H.~Ye, X.~Peng, B.~Zhao, and X.~Qu, ``Accelerated {MRI} reconstruction with separable and enhanced low-rank {H}ankel regularization,'' \emph{IEEE Trans. Med. Imaging}, vol.~41, no.~9, pp. 2486--2498, 2022.

\bibitem{29}
S.~Sun and Y.~D. Zhang, ``4{D} automotive radar sensing for autonomous vehicles: A sparsity-oriented approach,'' \emph{IEEE J. Sel. Topics Signal Process.}, vol.~15, no.~4, pp. 879--891, 2021.

\bibitem{30}
M.~Zhang, Y.~Liu, H.~Zhang, and Y.~Chen, ``Incoherent noise suppression of seismic data based on robust low-rank approximation,'' \emph{IEEE Trans. Geosci. Remote Sens.}, vol.~58, no.~12, pp. 8874--8887, 2020.

\bibitem{41}
Q.~Yang, Z.~Wang, K.~Guo, C.~Cai, and X.~Qu, ``Physics-driven synthetic data learning for biomedical magnetic resonance: The imaging physics-based data synthesis paradigm for artificial intelligence,'' \emph{IEEE Signal Process. Mag.}, vol.~40, no.~2, pp. 129--140, 2023.

\bibitem{31}
M.~Mayzel, K.~Kazimierczuk, and V.~Y. Orekhov, ``The causality principle in the reconstruction of sparse {NMR} spectra,'' \emph{Chem. Commun.}, vol.~50, no.~64, pp. 8947--8950, 2014.

\bibitem{32}
J.-F. Cai, E.~J. Cand{\`e}s, and Z.~Shen, ``A singular value thresholding algorithm for matrix completion,'' \emph{SIAM J. Optimiz.}, vol.~20, no.~4, pp. 1956--1982, 2010.

\bibitem{33}
Y.~Hu, D.~Zhang, J.~Ye, X.~Li, and X.~He, ``Fast and accurate matrix completion via truncated nuclear norm regularization,'' \emph{IEEE Trans. Pattern Anal. Mach. Intell.}, vol.~35, no.~9, pp. 2117--2130, 2012.

\bibitem{34}
S.~Boyd, N.~Parikh, and E.~Chu, ``Distributed optimization and statistical learning via the alternating direction method of multipliers,'' \emph{Found. Trends Mach. Learn.}, 2011.

\bibitem{42}
Y.~Zhou, C.~Qian, Y.~Guo, Z.~Wang, J.~Wang, B.~Qu, D.~Guo, Y.~You, and X.~Qu, ``Xcloud-p{FISTA}: {A} medical intelligence cloud for accelerated {MRI},'' in \emph{2021 43rd Annual International Conference of the IEEE Engineering in Medicine \& Biology Society (EMBC)}.\hskip 1em plus 0.5em minus 0.4em\relax IEEE, 2021, pp. 3289--3292.

\bibitem{35}
Z.~Tu, Z.~Wang, J.~Zhan, Y.~Huang, X.~Du, M.~Xiao, X.~Qu, and D.~Guo, ``A partial sum of singular-value-based reconstruction method for non-uniformly sampled {NMR} spectroscopy,'' \emph{IET Signal Proc.}, vol.~15, no.~1, pp. 14--27, 2021.

\bibitem{36}
C.~P. Butts, C.~R. Jones, E.~C. Towers, J.~L. Flynn, L.~Appleby, and N.~J. Barron, ``Interproton distance determinations by {NOE}--surprising accuracy and precision in a rigid organic molecule,'' \emph{Org. Biomol. Chem}, vol.~9, no.~1, pp. 177--184, 2011.

\bibitem{37}
R.~Dass, P.~Kasprzak, W.~Ko{\'z}mi{\'n}ski, and K.~Kazimierczuk, ``Artifacts in time-resolved {NUS}: A case study of {NOE} build-up curves from {2D NOESY},'' \emph{J. Magn. Reson.}, vol. 265, pp. 108--116, 2016.

\bibitem{38}
K.~Hu, W.~M. Westler, and J.~L. Markley, ``Simultaneous quantification and identification of individual chemicals in metabolite mixtures by two-dimensional extrapolated time-zero ${}^{\text{1}}\text{H-}{}^{\text{13}}\text{C}$ {HSQC} ({${\text{HSQC}_{0}}$)},'' \emph{J. Am. Chem. Soc.}, vol. 133, no.~6, pp. 1662--1665, 2011.

\bibitem{39}
K.~Hu, J.~J. Ellinger, R.~A. Chylla, and J.~L. Markley, ``Measurement of absolute concentrations of individual compounds in metabolite mixtures by gradient-selective time-zero ${}^{\text{1}}\text{H-}{}^{\text{13}}\text{C}$ {HSQC} with two concentration references and fast maximum likelihood reconstruction analysis,'' \emph{Anal. Chem.}, vol.~83, no.~24, pp. 9352--9360, 2011.

\bibitem{51}
H.~W. Engl, ``Discrepancy principles for tikhonov regularization of ill-posed problems leading to optimal convergence rates,'' \emph{J. Optim. Theory Appl.}, vol.~52, pp. 209--215, 1987.

\bibitem{44}
T.~Qiu, W.~Liao, Y.~Huang, J.~Wu, D.~Guo, D.~Liu, X.~Wang, J.-F. Cai, B.~Hu, and X.~Qu, ``An automatic denoising method for {NMR} spectroscopy based on low-rank {H}ankel model,'' \emph{IEEE Trans. Instrum. Meas.}, vol.~70, pp. 1--12, 2021.

\bibitem{43}
J.~Wu, R.~Xu, Y.~Huang, J.~Zhan, Z.~Tu, X.~Qu, and D.~Guo, ``Fast {NMR} spectroscopy reconstruction with a sliding window based {H}ankel matrix,'' \emph{J. Magn. Reson.}, vol. 342, p. 107283, 2022.

\bibitem{48}
K.~Unterforsthuber and K.~Bergmann, ``Mathematical separation procedure of broadline proton {NMR} spectra of crystalline polymers into components,'' \emph{J. Magn. Reson.}, vol.~33, no.~3, pp. 483--495, 1979.

\bibitem{49}
Y.~Huang, J.~Zhao, Z.~Wang, V.~Orekhov, D.~Guo, and X.~Qu, ``Exponential signal reconstruction with deep {H}ankel matrix factorization,'' \emph{IEEE Trans. Neural Networks Learn. Syst.}, vol.~34, no.~9, pp. 6214--6226, 2023.

\bibitem{50}
Z.~Wang, D.~Guo, Z.~Tu, Y.~Huang, Y.~Zhou, J.~Wang, L.~Feng, D.~Lin, Y.~You, T.~Agback, V.~Orekhov, and X.~Qu, ``A sparse model-inspired deep thresholding network for exponential signal reconstruction—application in fast biological spectroscopy,'' \emph{IEEE Trans. Neural Networks Learn. Syst.}, vol.~34, no.~10, pp. 7578--7592, 2023.

\end{thebibliography}

\end{document}